\documentclass[5p, twocolumn, 10pt, sort&compress]{elsarticle}
\journal{Computer Physics Communications}

\newcommand{\preprintmod}[2]{#2}

\usepackage{graphicx}
\usepackage[hidelinks, colorlinks]{hyperref}
\hypersetup{citecolor=blue}
\usepackage[utf8]{inputenc}
\usepackage[T1]{fontenc}
\usepackage[english]{babel}
\usepackage{ragged2e}
\usepackage{amsfonts, amsmath, amsthm, amssymb, bm, bbm, mathtools, derivative}
\usepackage{orcidlink}
\usepackage{indentfirst}

\usepackage[dvipsnames]{xcolor}
\usepackage[most]{tcolorbox}
\usepackage{listings}
\makeatletter
\newlength{\aboveskipoflstlisting}
\makeatother
\newtcbox{\pyinline}{
on line,
enhanced jigsaw,
breakable,
nobeforeafter,
tcbox raise base,
colback=blue!10,
boxrule=0pt,
boxsep=0pt,
arc=0pt,
left=2pt,
right=2pt,
top=2pt,
bottom=2pt,
fontupper=\ttfamily,
}
\newtcolorbox{SyntaxBlock}{
breakable,
colback=yellow!10,
boxrule=\fboxrule,
arc=0pt,
left=3pt,
right=3pt,
top=3pt,
bottom=3pt,
fontupper=\ttfamily,
before skip=0.5em,
after skip=0.5em,
after=\noindent\ignorespaces
}
\newtcolorbox{CodeBlock}{
breakable,
colback=blue!10,
boxrule=\fboxrule,
arc=0pt,
left=3pt,
right=3pt,
top=3pt,
bottom=3pt,
fontupper=\ttfamily,
before skip=0.5em,
after skip=0.5em,
after=\noindent\ignorespaces,
before upper={
    \obeylines
    \setlength{\parindent}{0pt}
    \everypar{%
      \hangindent=1.5em
      \hangafter=1
      \makebox[1.5em][l]{\textgreater}%
    }
  }
}
\newtcolorbox{RenderBlockColorBox}{
breakable,
colback=red!10,
boxrule=\fboxrule,
arc=0pt,
left=3pt,
right=3pt,
top=2pt,
before upper = {%
    \fcolorbox{red}{red!10}{\textcolor{red}{\texttt{\LaTeX~RENDER}}}
    \vspace*{-\baselineskip}
},
before skip=0.5em,
after skip=0.5em,
after=\noindent\ignorespaces
}
\newcommand{\RenderBlock}[1]{
\vspace*{-1.5\aboveskipoflstlisting}
\begin{RenderBlockColorBox}
\begin{align*}
#1
\end{align*}
\end{RenderBlockColorBox}
}
\newcommand{\symqups}{\texttt{SymQuPS}}
\newcommand{\sympy}{{SymPy}}

\newcommand{\aop}{\hat{a}}

\newcommand{\adop}{\aop^\dagger}

\newcommand{\PT}[1]{\left({#1}\right)}
\newcommand{\SB}[1]{\left[{#1}\right]}
\newcommand{\CB}[1]{\left\{{#1}\right\}}
\newcommand{\VB}[1]{\left|{#1}\right|}

\newcommand{\CG}[1]{\mathcal{W}_s\left[{#1}\right]}
\newcommand{\iCG}[1]{\mathcal{W}_s^{-1}\left[{#1}\right]}
\newcommand{\dStar}{\mathbin{\widehat{\star}_s}}
\newcommand{\aopadop}{\left(\aop,\adop\right)}
\newcommand{\alphacoord}{\left(\alpha,\alpha^*\right)}

\newcommand{\Star}{\mathbin{\star_s}}

\newcommand{\pdiff}[1]{\partial_{#1}}

\newcommand{\DRdiff}[1]{\pdiff{#1}^{R,\diamond}}

\newcommand{\Fop}{\hat{F}}
\newcommand{\Gop}{\hat{G}}

\newcommand{\trace}[1]{\mathrm{tr}\left({#1}\right)}

\newcommand{\ket}[1]{\left|{#1}\right\rangle}
\newcommand{\bra}[1]{\left\langle{#1}\right|}

\usepackage{tabularx, seqsplit}

\begin{document}\sloppy

\begin{frontmatter}

\title{SymQuPS: Symbolic Quantum Phase Space Algebra in Python}

\author[a]{Hendry M. Lim\corref{author}\fnref{currentaddress}}
\ead{hendry.minfui.lim@u.nus.edu}

\author[a]{Donny Dwiputra}
\author[a,b]{Ahmad R. T. Nugraha}

\cortext[author] {Corresponding author.}

\fntext[currentaddress]{\textit{Current address:} Department of Materials Science and Engineering, National University of Singapore, Singapore 117575, Republic of Singapore; and Center for Quantum Technologies, National University of Singapore, Singapore 117543, Republic of Singapore.}

\address[a]{Research Center for Quantum Physics, National Research and Innovation Agency (BRIN), South Tangerang 15314, Indonesia}
\address[b]{Department of Engineering Physics, School of Electrical Engineering, Telkom University, Bandung 40257, Indonesia}


\begin{abstract}
We present \texttt{SymQuPS}, a SymPy-based algebra system in Python mainly aimed toward the phase space representation of quantum mechanics within the Cahill-Glauber formalism (including the Glauber-Sudarshan $P$, Wigner, and Husimi $Q$ representations). By extension, the package serves as an algebraic venue for canonical quantization. A key feature is the phase space representation of an arbitrary Lindblad master equation, which gives the phase space equation of motion of the quantum system. We describe the core functionalities of the package, consisting of $s$-ordered operators, the star products, and the phase space representation. Some examples of use are given to illustrate the application of the package, and the package's performance in typical use cases is discussed.
\end{abstract}


\begin{keyword}
quantum phase space \sep Cahill-Glauber \sep Moyal star product \sep Wigner function \sep Glauber-Sudarshan P representation \sep Husimi Q function \sep canonical quantization

\end{keyword}

\end{frontmatter}


\noindent {\bf PROGRAM SUMMARY}

\begin{small}
\noindent
{\em Program Title:} \texttt{SymQuPS}
\\
{\em Developer's repository link:} \url{https://github.com/hendry24/SymQuPS}
\\
{\em Licensing provisions:} Apache License 2.0
\\
{\em Programming language:} Python
\\
{\em Nature of problem:} Algebraic evaluations of the phase space representation of a quantum mechanical operator in the Cahill-Glauber formalism (which includes the Wigner, Glauber-Sudarshan, and Husimi representation) and, by extension, canonical quantization of classical observables. 
\\
{\em Solution method:} Bopp shift evaluations of star products and hatted star products, implemented fully using SymPy, a computer algebra system library in Python. 
\end{small}

\section{Introduction}

There are several equally valid formulations of quantum mechanics~\cite{styer_nine_2002}, among which the phase space formulation is arguably less well known. Its obscurity may be attributed to two things: (i)~the development of the celebrated path integral formalism by Feynman~\cite{Feynman1948}, just several years after a complete formulation (independently done) by Groenewold~\cite{Groenewold1946} and Moyal~\cite{Moyal1949}, following the work of Wigner~\cite{Wigner1932}, and (ii)~how the problems tackled by the formulation are, by no means, impractically harder when tackled by the other formulations.  Historically speaking, it is interesting to note that only near the end of the twentieth century did the phase space formulation gain prominence~\cite{ZachosOnWhyThePhaseSpaceFormalismIsObscure}, mainly due to its possibility of observing dynamics that may not be as intuitively understood through the other formulations, such as decoherence and the classical limit of a quantum system~\cite{Case2008, curtright_concise_2013}. The phase space formalism has been used intensively in quantum optics, where, for example, Gaussian states of light exhibit interesting properties~\cite{Hillery1984Distribution, schleich2011quantum, leonhardt_measuring_1997, curtright_concise_2013, Buzek1995-fm}.  Recently, it has also found uses in numerous other topics, such as plasma physics and mesoscopic systems (for more examples, see the Introduction section of Ref.~\cite{curtright_concise_2013}).

A hallmark of the quantum phase space description is the \emph{star product}, which encodes the noncommutative nature of quantum mechanics into the space of scalar functions. Unfortunately, this additional noncommutative structure results in the phase space formulation being computationally costly. Numerical methods aside, analytic calculations involving the star product quickly grow in complexity as the number of quantum objects involved increases. Here, we use ``quantum object'' as an umbrella term for quantum states and their transformations. This increase in complexity leads to cumbersome and error-prone algebra that discourages us from unraveling interesting dynamics presented by the resulting equation of motion. Therefore, a tool to evaluate complex star products is highly desirable.

As an anecdote, one of the authors of this work (H.M.L.) had to calculate a chain of four star products involving arguably simple expressions in the (q,p) phase space (see Section~\ref{section2} for the formalism),
\begin{equation*}
(q+ip)\star(q+ip)\star f(q,p)\star (q-ip)\star (q-ip).
\end{equation*}
The calculation by hand took approximately two hours and turned out to be miscalculated somewhere, rendering the result unusable. H.M.L. is arguably not very skilled with complex algebra of this nature, but this work would not exist otherwise. 

In this paper, we present SymQuPS, a Python symbolic package for the quantum phase space algebra that is fully based on {SymPy}~\cite{sympy}. Its core working principle is based on the ``Bopp shift''~\cite{Hillery1984Distribution} evaluation of the star product, which allows the user to symbolically evaluate the star product given that one factor is a power series in the phase space variables. In its current version, the package implements the phase space formulation in the Cahill-Glauber formalism~\cite{Cahill1969, Cahill1969.II}, which includes the paradigmatic Wigner $W$~\cite{Wigner1932}, Husimi $Q$~\cite{Husimi1940}, and Glauber-Sudarshan $P$~\cite{Glauber1963, Sudarshan1963} transforms. By extension, the package can attempt to evaluate the phase space representation of any Hilbert space function in the operators representing the phase space variables ($\hat{q},\hat{p},\aop,\adop$). This feature, in turn, allows the user to attempt to transform any infinite-dimensional (or continuous variable) equation of motion, generally in the Lindblad form, into the corresponding equation of motion for the phase space representation. These transforms are invertible, and their inverse implementation doubles as a quantization framework with a similar capability based on a recent finding~\cite{lim2025algebraicmachineryquantization}.

The rest of this paper is organized as follows. In Section~\ref{section2}, we briefly discuss the Cahill-Glauber formalism.  We then describe the package's workflow in Section~\ref{section3}, give some examples of use in Section~\ref{section4}, and discuss the package's performance in Section~\ref{section5}. Finally, Section~\ref{section6} concludes the paper. 


\section{Mathematical Formalism}\label{section2}

This section provides a brief overview of the Cahill-Glauber formalism, the star product, and the time evolution of a quantum state. Readers already familiar with these topics may skip ahead to the next section. We refer mainly to Refs.~\cite{Hillery1984Distribution, schleich2011quantum, leonhardt_measuring_1997, gerry_introductory_2008, curtright_concise_2013, Case2008, Soloviev2015, lim_transient_2025, Buzek1995-fm, Cahill1969, Cahill1969.II} to write this section. Although the package supports a multipartite description, we shall discuss the formalism for a unipartite system, for simplicity. The multipartite version for relevant formulae is given in Appendix~\ref{appsec:multipartite}.

\subsection{The Cahill-Glauber correspondence} \label{subsec:CG_correspondence}

The discrepancy between the commutativity of phase space variables and the noncommutativity of their operator counterparts raises an \emph{ambiguity} in the connection between the two. Readers familiar with quantization will recognize this ambiguity as the ordering problem. Meanwhile, those working with phase space representations may relate the ambiguity to the non-uniqueness of the phase space representation. In fact, quantization and phase space representation are part of a correspondence: for every quantization map, there exists a corresponding phase space representation. 

One family of such correspondence is the Cahill-Glauber correspondence. The Cahill-Glauber transform can be seen as the interpolation among the ``big three'' of phase space representations of quantum mechanics: the paradigmatic Wigner function~\cite{Wigner1932}, the Glauber-Sudarshan $P$ distribution~\cite{Glauber1963, Sudarshan1963}, and the Husimi Q function~\cite{Husimi1940}. The inverses of these transformations are the generalized recipe for quantizing a classical system under the ``big three'' operator ordering choices: the Weyl or symmetric ordering, the normal ordering, and the antinormal ordering, respectively~\cite{hall_2013_QuantumTheoryMathematicians}. More formally, the CG correspondence establishes a connection between the $\alphacoord$ phase space, which is related to the usual $\PT{q,p}$ phase space by\footnote{One may question the decision to include the quantum constant $\hbar$ in a classical phase space. This form of phase space variables is particularly useful as it greatly simplifies the mathematical expressions within the formulation, and $\hbar$ can be treated as nothing more than a number. For the curious, the Poisson bracket within this phase space reads
\begin{equation}
    \CB{f,g}_\mathrm{PB} = \frac{1}{i\hbar} \PT{\pdv{f}{\alpha}\pdv{g}{\alpha^*}-\pdv{g}{\alpha}\pdv{f}{\alpha^*}}.
\end{equation}
}
\begin{equation}\label{eq:alpha_definition}
    \alpha = \frac{\zeta q+i\zeta^{-1} p}{\sqrt{2\hbar}},
\end{equation}
where $\zeta$ is a positive real parameter, e.g., $\zeta=\sqrt{m\omega}$ for a simple harmonic oscillator under Hooke's law. Within this phase space, $\alpha$ and $\alpha^*$ are treated as independent variables for all analytical purposes, such as differentiation (e.g., $\odv{\alpha}/{\alpha^*}=0$). The corresponding annihilation and creation operators (collectively called the ``ladder operators'') are given by
\begin{equation}
    \hat{a} = \frac{\zeta\hat{q}+i\zeta^{-1} \hat{p}}{\sqrt{2\hbar}}
\end{equation}
The Cahill-Glauber (CG) transform of some Hilbert space operator $\hat{F}$ is given by
\begin{equation}
    f^{(s)}\alphacoord = \CG{\hat{F}}\PT{\alpha,\alpha^*} = \mathrm{Tr}\PT{\hat{F}{\mathcal{T}}_{-s}},
\end{equation}
where the CG kernel is given by
\begin{equation}\label{eq:CG_kernel}
    \mathcal{T}_{s}\alphacoord = \int_{\mathbb{R}^2} \frac{\odif[order=2]{\beta}}{\pi} \hat{D}\PT{\beta} \exp\PT{\frac{s}{2}\VB{\beta}^2 + \alpha\beta^* - \alpha^*\beta} ,
\end{equation}
where $\hat{D}\PT{\beta}=e^{\beta\adop-\beta^*\aop}$ is known as the displacement operator. The parameter $s$ can formally take on any complex value, but the physically relevant one lies on the real number line between $-1$ and $1$ (inclusive). The inverse Cahill-Glauber (iCG) transform of some phase space function $f=f\alphacoord$ is given by
\begin{equation}
    \hat{F}^{(s)} = \iCG{f} = \int_{\mathbb{R}^2} \frac{\odif[order=2]{\alpha}}{\pi} f\mathcal{T}_s .
\end{equation}
Some notable properties:
\begin{itemize}
    \item Let $\CB{\PT{\adop}^m\aop^n}_s$ be the $s$-ordering of $\PT{\adop}^m\aop^n$. It does not matter what ordering is written inside the $s$-ordering bracket; only the number of $\aop$ and $\adop$ matters. For example, the antinormal ordering is given by $\CB{\PT{\adop}^m\aop^n}_{-1}=\aop^n\PT{\adop}^m$. Different orderings are related by
    \begin{equation}
    \begin{split}
        \preprintmod{
        \CB{\PT{\adop}^m\aop^n}_s = \sum_{k=0}^{\min\PT{m,n}} k! \binom{m}{k} \binom{n}{k} \PT{\frac{t-s}{2}}^{k} \CB{\PT{\adop}^{m-k}\aop^{n-k}}_t.
        }{
        &\CB{\PT{\adop}^m\aop^n}_s \\
        &\quad = \sum_{k=0}^{\min\PT{m,n}} k! \binom{m}{k} \binom{n}{k} \PT{\frac{t-s}{2}}^{k} \CB{\PT{\adop}^{m-k}\aop^{n-k}}_t.
        }
    \end{split}
    \end{equation}
    The $s$-parameterized CG transform of an $s$-ordering is equal to the operator-to-scalar substitution of its bracket content,
    \begin{equation}
        \CG{\CB{\PT{\adop}^m\aop^n}_s} = \PT{\alpha^*}^m \alpha^n.
    \end{equation}
    Its inverse is better-known: the $s$-ordered quantization of a monomial in the phase space variables is equal to the $s$-ordering of the monomial's scalar-to-operator substitution (or promotion). For example, we can quantize $\alpha^*\alpha$ into $\adop\aop$ for the normal-ordered ($s=1$) ordering choice, or into $\PT{\aop\adop+\adop\aop}/2$ for the Weyl-ordered ($s=0$) ordering choice.
    \item The CG transform of the density matrix gives the phase space state function $W_{-s}$ of the quantum state,
    \begin{equation}
        \CG{\rho} = W_{-s}
    \end{equation}
    For $s=-1,0,1$, we respectively have the Glauber-Sudarshan $\PT{\pi P}$ representation\footnote{The factor $\pi$ comes from the definition of the representation,
    \begin{equation}
    \begin{split}
        \rho &= \int_{\mathbb{R}^2} \odif[order=2]{\alpha} P\alphacoord \ket{a}\bra{a}
        = \int_{\mathbb{R}^2} \frac{\odif[order=2]{\alpha}}{\pi} \SB{\pi P\alphacoord} \ket{a}\bra{a},
    \end{split}
    \end{equation}
    as compared to the CG kernel given in Eq.~\eqref{eq:CG_kernel}.}, the Wigner function $W$, and the Husimi $Q$ function. The sign flip reflects the optical equivalence theorem,
    \begin{equation}
    \begin{split}
        \trace{\rho \left\{\PT{\adop}^m\aop^n\right\}_t} 
        &=
        \int_{\mathbb{R}^2} \frac{\odif[order=2]{\alpha}}{\pi} \CG{\left\{\PT{\adop}^m\aop^n\right\}_t} \mathcal{W}_{-t}\SB{\rho} 
    \\
        &=
        \int_{\mathbb{R}^2} \frac{\odif[order=2]{\alpha}}{\pi} \PT{\alpha^*}^m\alpha^n W_t\alphacoord.
    \end{split}
    \end{equation}
    \item The CG transform and its inverse are complex linear. That is,
        \begin{equation}
            \CG{\alpha \Fop + \beta \Gop} = \alpha f^{(s)} + \beta g^{(s)}, \quad \alpha,\beta \in \mathbb{C},
        \end{equation}
    and likewise for the iCG transform.
    \item For real $s$, the CG transform of a Hermitian conjugate of an operator is equal to the complex conjugate of the CG transform of that operator,
    \begin{equation}
        \CG{\Fop^\dagger} = \SB{f^{(s)}}^*,\quad s\in \mathbb{R}.
    \end{equation}
    On the flip side, the iCG transform of the complex conjugate of a function is equal to the Hermitian conjugate of the iCG transform of that function, as evident by taking the iCG transform of the above equation.
    \item The CG transform of a function in $\aop$ only is the same function with $\aop$ replaced by $\alpha$, and similarly for $\adop$,
    \begin{equation}\label{eq:CG_of_single_variable_function}
        \CG{f\PT{\aop}} = f\PT{\alpha}.
    \end{equation}
    The inverse of this property applies, as can be seen by taking the iCG transform of both sides of Eq.~\eqref{eq:CG_of_single_variable_function}.
    \item The iCG transform of derivatives with respect to $\alpha$ and $\alpha^*$ are given by
    \begin{equation}
        \iCG{\pdv{f}{\alpha}} = \pdv{\hat{F}^{s}}{\aop}, \quad \iCG{\pdv{f}{\alpha^*}} = \pdv{\hat{F}^{s}}{\adop}, 
    \end{equation}
    where the formal derivatives with respect to the ladder operators are defined as $\partial_{\aop}\hat{F} = -\SB{\adop,\hat{F}}$ and $\partial_{\adop}\Fop = \SB{\aop,\Fop}$, e.g., $\partial_{\aop}\aop = 1$ and $\partial_{\adop} \PT{\adop}^2 = 2\adop$. It is worth noting that it is thanks to these formal derivatives that we can conveniently write the relations found throughout this text. The situation is more complicated had we used the $\PT{q,p}$ phase space and the $\PT{\hat{q},\hat{p}}$ operator basis.
    %
    %
    %
\end{itemize}

\subsection{The star products}

The CG transform of a product is given by the star product between the CG transforms of its factors,
\begin{equation}
    \CG{\Fop\Gop} = f^{(s)}\Star g^{(s)}.
\end{equation}
Formally, the star product takes the form of an integral. However, for practical purposes, it suffices to consider the better-known differential form,
\begin{equation}\label{eq:Star}
    \Star = \exp\PT{ \frac{s+1}{2} \overset{\leftarrow}{\partial}_\alpha \overset{\rightarrow}{\partial}_{\alpha^*} + \frac{s-1}{2} \overset{\leftarrow}{\partial}_{\alpha^*} \overset{\rightarrow}{\partial}_\alpha }.
\end{equation}
The arrows on top of the partial differentiation operators indicate the direction in which they operate. These directed derivatives treat the variables belonging to the same expression as them as constants. While this property is not evident using Eq.~\eqref{eq:Star}, we can consider one useful evaluation method called the Bopp shift: since $\exp\PT{A\partial_x}f(x)=f(x+A)$ given that $A$ is independent of $x$, we may write
\begin{equation}
\begin{split}
    \preprintmod{
        f\alphacoord \Star g\alphacoord &= f\PT{\alpha+\frac{s+1}{2}\overset{\rightarrow}{\partial}_{\alpha^*}, \alpha^*+\frac{s-1}{2} \overset{\rightarrow}{\partial}_\alpha} g\alphacoord
    \\
    &= f\PT{\alpha+\frac{s+1}{2}{\partial}_{\beta^*}, \alpha^*+\frac{s-1}{2} {\partial}_\beta} g\PT{\beta,\beta^*} \Bigg|_{\beta\mapsto\alpha}
    }{
        &f\alphacoord \Star g\alphacoord 
    \\
    &\quad = f\PT{\alpha+\frac{s+1}{2}\overset{\rightarrow}{\partial}_{\alpha^*}, \alpha^*+\frac{s-1}{2} \overset{\rightarrow}{\partial}_\alpha} g\alphacoord
    \\
    &\quad = f\PT{\alpha+\frac{s+1}{2}{\partial}_{\beta^*}, \alpha^*+\frac{s-1}{2} {\partial}_\beta} g\PT{\beta,\beta^*} \Bigg|_{\beta\mapsto\alpha}
    }
\end{split}
\end{equation}
for any two phase space functions $f$ and $g$ (not necessarily a CG transformation result, hence no $(s)$ superscript). The second line of this equation illustrates the property of the directed derivatives mentioned above. The Bopp shift can also be applied in the right-hand direction, in which case we have
\begin{equation}
\begin{split}
    \preprintmod{
        f\alphacoord \Star g\alphacoord &= f\alphacoord g\PT{\alpha+\frac{s-1}{2}\overset{\leftarrow}{\partial}_{\alpha^*}, \alpha^*+\frac{s+1}{2} \overset{\leftarrow}{\partial}_\alpha}
    \\
    &= g\PT{\alpha+\frac{s-1}{2}{\partial}_{\beta^*}, \alpha^*+\frac{s+1}{2} {\partial}_\beta} f\PT{\beta,\beta^*} \Bigg|_{\beta\mapsto\alpha}
    }{
        &f\alphacoord \Star g\alphacoord 
        \\ &\quad = f\alphacoord g\PT{\alpha+\frac{s-1}{2}\overset{\leftarrow}{\partial}_{\alpha^*}, \alpha^*+\frac{s+1}{2} \overset{\leftarrow}{\partial}_\alpha}
    \\
    &\quad = g\PT{\alpha+\frac{s-1}{2}{\partial}_{\beta^*}, \alpha^*+\frac{s+1}{2} {\partial}_\beta} f\PT{\beta,\beta^*} \Bigg|_{\beta\mapsto\alpha}
    }
\end{split}
\end{equation}
The star products are noncommutative, becoming the hallmark of quantum mechanics in phase space. In particular, the CG transform of $1/(i\hbar)$ times the commutator is given by
\begin{equation}\label{eq:CG_bracket}
\begin{split}
    \CG{\frac{1}{i\hbar}\SB{\Fop,\Gop}} &= \frac{1}{i\hbar} \SB{ f^{(s)} \Star g^{(s)} - g^{(s)} \Star f^{(s)} }
    \\
    &= \CB{\CB{ f^{(s)}, g^{(s)} }}_s,
\end{split}
\end{equation}
which gives the paradigmatic Moyal bracket for $s=0$. We can additionally define the ``phase space Bopp operators'' (PSBOs),
\begin{equation}
\begin{split}
    \overset{\leftarrow}{\mathcal{B}}_\alpha &= \alpha + \frac{s-1}{2} \overset{\leftarrow}{\partial}_{\alpha^*}, \\
    \overset{\leftarrow}{\mathcal{B}}_{\alpha^*} &= \alpha^* + \frac{s+1}{2} \overset{\leftarrow}{\partial}_{\alpha}, \\ 
    \overset{\rightarrow}{\mathcal{B}}_\alpha &= \alpha + \frac{s+1}{2} \overset{\rightarrow}{\partial}_{\alpha^*}, \\
    \overset{\rightarrow}{\mathcal{B}}_{\alpha^*} &= \alpha^* + \frac{s-1}{2} \overset{\rightarrow}{\partial}_{\alpha},
\end{split}
\end{equation}
to write the properties
\begin{equation}
\begin{split}
    f^{(s)}\star g = \Fop\Big|_{\aop\mapsto \overset{\rightarrow}{\mathcal{B}}_\alpha, \adop \mapsto \overset{\rightarrow}{\mathcal{B}}_{\alpha^*}} g, \\
    f\star g^{(s)} = f \Gop\Big|_{{\aop\mapsto \overset{\leftarrow}{\mathcal{B}}_\alpha, \adop \mapsto \overset{\leftarrow}{\mathcal{B}}_{\alpha^*}}}.
\end{split}
\end{equation}
We note again that functions without the superscript ``$(s)$'' denote generic functions in the corresponding space, which may or may not be a result of a CG/iCG transform.

\subsection{The hatted star products}

The iCG transform of a product is given by the hatted star product between the iCG transforms of its factors,
\begin{equation}
    \iCG{fg} = \Fop^{(s)} \dStar \Gop^{(s)}.
\end{equation}
Like its phase space counterpart, this mappping is formally an integral. Practically, however, we can consider the differential form
\begin{equation}
    \dStar = \exp\PT{-\frac{s+1}{2}\overset{\leftarrow}{\partial}_{\aop} \overset{\rightarrow}{\partial}_{\adop}-\frac{s-1}{2}\overset{\leftarrow}{\partial}_{\adop} \overset{\rightarrow}{\partial}_{\aop}},
\end{equation}
where the directional derivatives apply thanks to the commutation property $\SB{\hat{A},\hat{B}\hat{C}} = \SB{\hat{A},\hat{B}}\hat{C} + \hat{B}\SB{\hat{A},\hat{C}}$, and its variations. The Bopp shift similarly applies, and we can write
\begin{equation}
\begin{split}
    \preprintmod{
            \Fop\aopadop \dStar \Gop\aopadop 
    &=
    \Fop\PT{\aop-\frac{s+1}{2} \overset{\rightarrow}\partial_{\adop}, \adop - \frac{s-1}{2} \overset{\rightarrow}{\partial}_{\aop} } \Gop \aopadop
    \\
    &= 
    \Fop\PT{\aop-\frac{s+1}{2} \partial_{\hat{b}^\dagger}, \adop - \frac{s-1}{2} {\partial}_{\hat{b}} } \Gop \PT{\hat{b}, \hat{b}^\dagger} \Bigg|_{\hat{b} \mapsto \aop},
    }{
            &\Fop\aopadop \dStar \Gop\aopadop 
    \\&\quad =
    \Fop\PT{\aop-\frac{s+1}{2} \overset{\rightarrow}\partial_{\adop}, \adop - \frac{s-1}{2} \overset{\rightarrow}{\partial}_{\aop} } \Gop \aopadop
    \\
    \\&\quad = 
    \Fop\PT{\aop-\frac{s+1}{2} \partial_{\hat{b}^\dagger}, \adop - \frac{s-1}{2} {\partial}_{\hat{b}} } \Gop \PT{\hat{b}, \hat{b}^\dagger} \Bigg|_{\hat{b} \mapsto \aop},
    }
\end{split} 
\end{equation}
and likewise for Bopp shifting the right-hand-side argument. Similar to the PSBOs, here we have the ``Hilbert space Bopp superoperators'' (HSBSs),
\begin{equation}\label{eq:HSBS}
\begin{split}
    \overset{\leftarrow}{\mathcal{B}}_{\aop} &= \aop - \frac{s-1}{2} \overset{\leftarrow}{\partial}_{\adop}, \\
    \overset{\leftarrow}{\mathcal{B}}_{\adop} &= \adop - \frac{s+1}{2} \overset{\leftarrow}{\partial}_{\aop}, \\ 
    \overset{\rightarrow}{\mathcal{B}}_{\aop} &= \aop - \frac{s+1}{2} \overset{\rightarrow}{\partial}_{\adop}, \\
    \overset{\rightarrow}{\mathcal{B}}_{\adop} &= \adop - \frac{s-1}{2} \overset{\rightarrow}{\partial}_{\aop},
\end{split}
\end{equation}
from which we have
\begin{equation}
\begin{split}
    \Fop^{(s)}\dStar \Gop &= f\Big|_{\alpha\mapsto \overset{\rightarrow}{\mathcal{B}}_{\aop}, \alpha^*\mapsto \overset{\rightarrow}{\mathcal{B}}_{\adop}} \Gop, \\
    \Fop \dStar \Gop^{(s)} &= \Fop g\Big|_{\alpha\mapsto \overset{\leftarrow}{\mathcal{B}}_{\aop}, \alpha^*\mapsto \overset{\leftarrow}{\mathcal{B}}_{\adop}}.
\end{split}
\end{equation}
Unlike the PSBOs, the HSBSs commute with each other, so they can be used to speed up the evaluations of hatted star products.

\subsection{Equations of motion}

In the Hilbert space formalism, a widely used evolution framework for a generally open quantum system is given by the Gorini–Kossakowski–Sudarshan–Lindblad (GKSL) equation, or simply the Lindblad master equation~\cite{Lindblad1976, Gorini1976, schlosshauer2007decoherence, breuer2002theory},
\begin{equation}\label{Eq:lindblad_me}
    \odv{\rho}{t} = \frac{1}{i\hbar}\left[\hat{H},\rho\right] + \sum_{jk} \gamma_{jk} \mathcal{D}\left(\hat{F}_j, \hat{F}_k\right)\left[\rho\right],
\end{equation}
where $\hat{H}$ is the Hamiltonian. Meanwhile,
\begin{equation}
    \mathcal{D}\left(\hat{F}_j,\hat{F}_k\right)\left[\rho\right] = \hat{F}_j\rho\hat{F}_k^\dagger - \frac{1}{2}\left\{\hat{F}_k^\dagger\hat{F}_j,\rho\right\}
\end{equation}
is the Lindblad dissipator, where $\left\{\hat{A},\hat{B}\right\} = \hat{A}\hat{B} + \hat{B}\hat{A}$ is the anticommutator. The elements of the positive semidefinite matrix $\left\{\gamma_{jk}\right\}$ describe the processes' rates. Mathematically, this is the most general form of the dissipators allowed by the formalism. In practice, however, it is more common to encounter the dissipators in the diagonal form, i.e., with $\gamma_{j,k\neq j}=0$, at least to our observation. It is straightforward to evaluate the CG transform of the Lindblad master equation using the properties discussed above. The state function $W_{-s}$ evolves according to
\begin{equation}
    \pdv{W_{-s}}{t} = \left\{\left\{\mathcal{H}^{(s)} , W_{-s}\right\}\right\}_s + \sum_{jk}\gamma_{jk}\mathcal{D}_{\Star} \left(\hat{F}_j,\hat{F}_k\right)\left[W_{-s}\right],
\end{equation}
where $\mathcal{H}^{(s)}=\CG{\hat{H}}$ and we have defined the $\Star$-dissipator $\mathcal{D}_{\Star}$ as the CG transform of the Lindblad dissipator,
\begin{equation}\begin{split}
    \mathcal{D}_{\Star} \left(\hat{F}_j,\hat{F}_k\right)\left[W_{-s}\right]
    &= 
    \SB{\mathcal{F}_j^{(s)}} \Star \SB{W_{-s}} \Star \SB{\mathcal{F}_k^{(s)}}^* 
    \\ &\quad 
    - \frac{1}{2} \SB{\mathcal{F}_k^{(s)}}^* \Star \SB{\mathcal{F}_j^{(s)}} \Star \SB{W_{-s}}
    \\ &\quad
    - \frac{1}{2}\SB{W_{-s}} \Star \SB{\mathcal{F}_k^{(s)}}^* \Star \SB{\mathcal{F}_j^{(s)}}.
\end{split}\end{equation}


\section{Symbolic Implementation with SymPy}\label{section3}

The package \symqups~is built entirely on top of \sympy~\cite{sympy}, a versatile computer algebra system (CAS) in Python, to implement symbolic quantum phase space algebra within the Cahill-Glauber correspondence framework. A power user may import \symqups~in its entirety to make use of its full capabilities. An average user may instead utilize the subpackage \pyinline{symqups.simple} to use the minimal version of the package for unipartite descriptions, where the phase space formulation is arguably the most useful. Throughout this section, we describe the structure of \symqups~in general. The subpackage \pyinline{symqups.simple} is then used in the next section, to illustrate some simple examples of use. 

All the functionalities in the package accept expressions in $\PT{\hat{q},\hat{p}}$ and $\PT{q,p}$, which is then converted into $\aopadop$ and $\alphacoord$, respectively. All user-level functionalities of the package, including those not mentioned in the main text, are listed in Appendix~\ref{appsec:docs}. These functionalities include \pyinline{qp2alpha} and \pyinline{alpha2qp}, which can be used to convert one coordinate/operator system to the other. 

For the remainder of this text, we shall assume that the readers are already familiar with the fundamentals of Python. We further assume that both \symqups~and SymPy are imported as follows,
\begin{CodeBlock}
import symqups as sq

import sympy
\end{CodeBlock}

It is worth noting that \emph{none} of the functionalities in our package is directly compatible with the \pyinline{sympy.physics.quantum} subpackage; the two should not be used together without appropriate knowledge of how to manipulate the objects in one package into another. The user should not need to use anything outside of \sympy's core functionalities, as we have developed \symqups~to be as self-contained as possible.


\subsection{Primary objects}

The physical scalars and Hilbert space operators are implemented as separate classes of objects, both subclassing the 
\pyinline{Base} class. This class is essentially modified \pyinline{sympy.Symbol}, which represents symbols with caching, allowing the creation of multiple instances referring to the same object. The modification allows for assigning additional attributes to the objects such as the subcripts that denote which subsystem an expression refers to. Additionally, we have a separate class for \pyinline{Constant}'s that can be set by the user and the class \pyinline{sOrdering} to $s$-order a given expression in $\aopadop$. 

\subsubsection{Scalars}

The phase space variables \pyinline{q} ($q$), \pyinline{p} ($p$), \pyinline{alpha} ($\alpha$), and \pyinline{alphaD} ($\alpha^*$) are separate subclasses of \pyinline{Scalar}, an object class for scalars that subclasses \pyinline{Base}. The syntax used to instantiate a \pyinline{q} object is\footnote{We note that the render block does not always show what is displayed by a Jupyter Notebook cell. For example, when the last line of the code block is a \pyinline{set} of SymPy objects, then Jupyter Notebook usually returns \pyinline{\{item\_1, item\_2, ...\}} where each item is the Python representation \pyinline{repr} of the SymPy object.}
\begin{CodeBlock}
q\_j = sq.q(sub="j")

q\_j
\end{CodeBlock}
\RenderBlock{
q_{j}
}
and likewise for the others. The custom attribute \pyinline{.sub} (for ``subscript'') is pre-treated by the internal function \pyinline{treat\_sub}, allowing the user to be more lenient with the input and ensuring that \pyinline{.sub} is always a \pyinline{sympy.Symbol}. The package infers what the ``system'' is by storing all the created \pyinline{.sub}'s inside an internal cache object. This cache is used for some functionalities in the package, namely phase space-variable--to--operator conversion, $\alphacoord$-to-$\PT{q,p}$ conversion, and the list of variables for \pyinline{StateFunction}, the object representing the phase space representation $W_s$ of a quantum state. The user should not instantiate their own \pyinline{StateFunction} object, so it is not exposed in the package's top-level namespace. Instead, the user can retrieve the object by running
\begin{CodeBlock}
    sq.W
\end{CodeBlock}
\RenderBlock{
    W_s
}
Its arguments can be accessed using the \pyinline{.args} attribute, and is always automatically updated whenever a new \pyinline{sub} is added to the cache. Since our formulation is based on the $\alphacoord$ phase space, the variables contained in \pyinline{W} are \pyinline{alpha} and \pyinline{alphaD} for a given \pyinline{.sub}. Additionally, \pyinline{W} contains the time variable \pyinline{sq.t} by default, which is not a user-level variable. Unlike other base objects within the package, a \pyinline{StateFunction} is not a subclass of \pyinline{Base}, but a subclass of Sympy's \pyinline{sympy.Function}. This means that one can straightforwardly instantiate its derivative with respect to its arguments using \pyinline{sympy.Derivative}.

\subsubsection{Hilbert space operators}

The Hilbert space operators corresponding to the phase space variables are subclasses of \pyinline{Operator}, which is another subclass of \pyinline{Base}. To instantiate these objects, the user may write\footnote{We remind the readers that, while the operator $\hat{A}_j$ only transforms the $j$th subsystem, it acts on the composite system's Hilbert space $\mathcal{H}_\mathrm{system}=\bigotimes_{j=0}^{N-1}\mathcal{H}_j$. That is, 
\begin{equation*}
    \hat{A}_j = \bigotimes_{k=0}^{j-1} \hat{1} \otimes \hat{A} \otimes \bigotimes_{k=j+1}^{N-1} \hat{1},
\end{equation*}
where $\hat{1}$ is the identity operator and $\hat{A}$ is the operator that acts only on a single subsystem's Hilbert space. As this is rarely stated in the literature, we hope this note can clear some confusion for the readers when using a subsystem-indexed operator alongside, say, a bra or a ket belonging to the composite Hilbert space.
}
\begin{CodeBlock}
qOp\_j = sq.qOp(sub="j")

qOp\_j
\end{CodeBlock}
\RenderBlock{
\hat{q}_j
}
and likewise for \pyinline{pOp} ($\hat{p}$), \pyinline{annihilateOp} ($\aop$), and \pyinline{createOp} ($\adop$). The \pyinline{.sub} attribute is the same as that of the \pyinline{Scalar} objects, sharing the same cache within the package. As such, the state function \pyinline{W} also updates when a new \pyinline{.sub} is created via \pyinline{Operator}. An \pyinline{Operator} object has an additional method \pyinline{.dagger}, which returns the object's Hermitian conjugate. For example,
\begin{CodeBlock}
    aOp\_j = sq.annihilateOp(sub="j")
    
    aOp\_j.dagger()
\end{CodeBlock}
\RenderBlock{
    \hat{a}^{\dagger}_{j}
}
Since \pyinline{Base} is a subclass of \pyinline{sympy.Symbol}, it cannot hold any arguments (i.e., its \pyinline{.args} attribute is an empty set), so an \pyinline{Operator} is taken as time-independent by SymPy. However, we need the density operator $\rho$ to be time-dependent. The class \pyinline{TimeDependentOp} implements time-dependent operators, but only accepts an instance of \pyinline{densityOp} as other operators are time-independent in the formalism. Like \pyinline{StateFunction}, both \pyinline{densityOp} and \pyinline{TimeDependentOp} are not user-level. The instance representing $\rho(t)$ can be retrieved using
\begin{CodeBlock}
    sq.rho
\end{CodeBlock}
\RenderBlock{
    {\rho_{}}(t)
}
Since SymPy's own \pyinline{sympy.Derivative} class treats \symqups' objects as if they are \pyinline{sympy.Symbol}'s, we have forbidden \pyinline{Operator} instances from being input as differentiation variables. To implement formal derivatives with respect to $\aop$ and $\adop$, \symqups~implements its own \pyinline{Derivative} class. The syntax for \pyinline{Derivative} is identical \pyinline{sympy.Derivative}. Internally, the class evaluates the formal derivative by turning them into the corresponding \pyinline{Commutator} instances. The output will then either be the worked out derivative or a \pyinline{sympy.Derivative} instance, depending on whether there are any differentiation variables left. For example,
\begin{CodeBlock}
    sq.Derivative(rho, aOp\_j)
\end{CodeBlock}
\RenderBlock{
    -\SB{\adop_j, {\rho}(t)}
}
and
\begin{CodeBlock}
    sq.Derivative(sq.W, aOp\_j, sq.t)
\end{CodeBlock}
\RenderBlock{
    \frac{\partial}{\partial t_{}} \left[{\rho_{}}(t),\hat{a}^{\dagger}_{j}\right]
}
the latter being a bona fide \pyinline{sympy.Derivative} instance.

\subsubsection{Constants}

There are three configurable \pyinline{Constant}'s in the package, user-accessible as global variables: the reduced Planck's constant \pyinline{hbar} ($\hbar$), the scaling parameter \pyinline{zeta} ($\zeta$) in Eq.~\eqref{eq:alpha_definition}, and the Cahill-Glauber $s$ parameter \pyinline{s}. Their values are stored in the \pyinline{.val} attribute, which can be set by the user. The default values are stored as the \pyinline{.default\_value} attribute, with which the user may restore the original value for the \pyinline{Constant}'s, say by writing
\begin{CodeBlock}
    s.val = s.default\_value
\end{CodeBlock}
for the $s$ parameter. The package implements checks on the values set by the user. For the $s$ parameter, the package will warn the user if they set the value outside the $[-1,1]$ real-valued interval. For the other two, the package raises an error if their values are not real and positive. The current value of $s$ determines the parameter used by most functionalities in the package when they are called, and becomes the default parameter for some others.

\subsubsection{The $s$-ordering}

The $s$-ordering of a Hilbert space expression is implemented as \pyinline{sOrdering} by the package. The syntax is
\begin{SyntaxBlock}
    sOrdering(expr, s)    
\end{SyntaxBlock}
By default, the parameter \pyinline{s} is set to the value of the CG $s$ parameter object discussed above. Upon instantiation, the class checks if \pyinline{expr} contains any ordering ambiguity. If not, \pyinline{expr} is returned as is. Otherwise, an \pyinline{sOrdering} instance is returned. For example, with \pyinline{s=1}, we have
\begin{CodeBlock}
    aOp, adOp = sq.annihilateOp(), sq.createOp() 
    
    \# The default `sub` is "" which creates an empty symbol.
    
    normal\_ordered = sq.sOrdering(aOp*adOp, 1)

    normal\_ordered
\end{CodeBlock}
\RenderBlock{
    \left\{ \hat{a}^{\dagger}_{} \hat{a}_{} \right\}_{s=1}
}
An \pyinline{sOrdering} object has the method \pyinline{.explicit} to explicitly write the ordering for $s=-1,0,1$; otherwise, the method returns the same object. Using the variable above, we have
\begin{CodeBlock}
    normal\_ordered.explicit()
\end{CodeBlock}
\RenderBlock{
    \hat{a}^{\dagger}_{} \hat{a}_{}
}
Additionally, the object has the method \pyinline{.express(t, explicit)} to return its $t$-ordered equivalent, applying \pyinline{.explicit} if specified. For example, we can get the antinormal-ordered equivalent of \pyinline{normal\_ordered} as follows,
\begin{CodeBlock}
    normal\_ordered.express(-1, True)
\end{CodeBlock}
\RenderBlock{
    -1 + \hat{a}_{} \hat{a}^{\dagger}_{}
}
More generally, the user can call \pyinline{.explicit} and \pyinline{.express} for any expression containing \pyinline{sOrdering} using the functions
\begin{SyntaxBlock}
    explicit\_sOrdering(expr)
\end{SyntaxBlock}
and
\begin{SyntaxBlock}
    express\_sOrdering(expr, t, explicit)
\end{SyntaxBlock}


\subsection{Star products and hatted star products} \label{subsec_core_func_star_hatted_star}

Star products and hatted star products can be evaluated using \pyinline{Star} and \pyinline{HattedStar}, respectively. Their outputs are always given in terms of $\alphacoord$ and $\aopadop$, even when expressions containing $(q,p)$ and $\PT{\hat{q},\hat{p}}$ is input. The package uses the Bopp shift to evaluate products, assuming at least one argument is a polynomial in the ladder operators, by implementing the expressions given in Appendix~\ref{appsec:explicit_formulae}. The syntax is given by
\begin{SyntaxBlock}
    Star(f1, f2, ...)
\end{SyntaxBlock}
which evaluates $f_1\Star f_2\Star \dots$, and likewise for \pyinline{HattedStar}. If both arguments of a given $\Star$ are Bopp-shiftable, then the left-hand-side argument is Bopp-shifted. The star products are associative, so the package evaluates evaluable star products in sequence until none remain. If at least one $\Star$ remains, the class constructs a \pyinline{Star} product chain containing factors that are not Bopp-shiftable. These star-product factors are stored in the \pyinline{.args} attribute of the object. For example, we have
\begin{CodeBlock}
    a, ad = sq.alpha(), sq.alphaD() 
    
    f = sympy.Function("f")(a, ad)

    sq.Star(a, f)
\end{CodeBlock}
\RenderBlock{
    \frac{s \frac{\partial}{\partial \overline{\alpha}_{}} f{\left(\alpha_{},\overline{\alpha}_{} \right)}}{2} + \alpha_{} f{\left(\alpha_{},\overline{\alpha}_{} \right)} + \frac{\frac{\partial}{\partial \overline{\alpha}_{}} f{\left(\alpha_{},\overline{\alpha}_{} \right)}}{2}
}
and
\begin{CodeBlock}
    sq.Star(sympy.exp(a), f)
\end{CodeBlock}
\RenderBlock{
    \left({e^{\alpha_{}}}\right)\mathbin{\star_s}\left({f{\left(\alpha_{},\overline{\alpha}_{} \right)}}\right)
}
We note that SymPy renders complex conjugates as $\overline{z}$ instead of $z^*$. The working principle is identical for \pyinline{HattedStar}. As shown by the above example, the package does not account for non-polynomial Bopp-shiftable functions such as $e^{\alpha}$ (see Appendix~\ref{appsec:other_boppable}).


\subsection{The Cahill-Glauber transform and its inverse}\label{section3C}

As an algebra system, \symqups~does not analytically evaluate the CG transform and its inverse; our implementation is algebraic based on the properties listed in Section~\ref{subsec:CG_correspondence}. The CG transform of a given Hilbert space expressions in $\aopadop$ can be evaluated using \pyinline{CGTransform} with the syntax
\begin{SyntaxBlock}
    CGTransform(expr)
\end{SyntaxBlock}
and likewise for its inverse, \pyinline{iCGTransform}. Upon considering different algorithms (e.g., using the PSBOs), we find that it is efficient to abuse the commutativity of scalar functions to treat a product. For \pyinline{CGTransform}, the package first separates a term into its factors by whether they are polynomial. The CG transform of each polynomial factor is evaluated by rewriting it into its $s$-ordered equivalent, then taking the operator-to-scalar conversion of the ordering-bracket contents. The CG transform of each nonpolynomial factor may be evaluated to some phase space expression if the package knows how to (e.g., exponentials containing only $\aop$ or $\adop$), or to a generic \pyinline{CGTransform} instance otherwise (e.g., exponentials containing both $\aop$ and $\adop$). The \pyinline{Star} product of these transformed factors is then returned. The package additionally knows how to treat \pyinline{sympy.Equality}, \pyinline{sOrdering} (even those not having the same ordering parameter as the \pyinline{s} constant), \pyinline{HattedStar}, \pyinline{Commutator}, and \pyinline{iCGTransform}. For example, we have
\begin{CodeBlock}
    sq.CGTransform(aOp*adOp)
\end{CodeBlock}
\RenderBlock{
    \frac{s}{2} + \alpha_{} \overline{\alpha}_{} + \frac{1}{2}
}
and
\begin{CodeBlock}
    sq.CGTransform(sympy.exp(aOp))
\end{CodeBlock}
\RenderBlock{
    e^{\alpha_{}}
}
and
\begin{CodeBlock}
    sq.CGTransform(sympy.exp(aOp*adOp))
\end{CodeBlock}
\RenderBlock{
    \mathcal{W}_{s={s}}\left[{e^{\hat{a}_{} \hat{a}^{\dagger}_{}}}\right]
}
Meanwhile, \pyinline{iCGTransform} separates a given term into polynomial and nonpolynomial parts, then turn the polynomial part into HSBSs which is then applied to the \pyinline{iCGTransform} of the nonpolynomial part, which is either a clean expression or an unevaluable \pyinline{iCGTransform} instance. The explicit formula for the HSBS application is given in Appendix~\ref{appsec:explicit_formulae}. This class can handle \pyinline{sympy.Equality}, \pyinline{sympy.Derivative}, \pyinline{Star}, and \pyinline{CGTransform}.


\subsection{The Lindblad master equation}

The implementation for the Lindblad master equation is similar to that of \texttt{PyBoLaNO}~\cite{lim_2025_PyBoLaNOPythonSymbolic}. The syntax is given by
\begin{SyntaxBlock}
    LindbladMasterEquation(H, *dissipators)
\end{SyntaxBlock}
or \pyinline{LME}, for short. Here, \pyinline{H} is the Hamiltonian. Meanwhile, each of the \pyinline{dissipators} can either be an expression or a sequence of two or three entries. If an expression is given, then it is taken as $\sqrt{\gamma_{jj}}\hat{F}_j$ with $\Fop_j=\Fop_k$ in Eq.~\eqref{Eq:lindblad_me}. If a two-entry sequence is given, then the first entry is taken as $\gamma_{jj}$ and the second entry is taken as $\hat{F}_j=\hat{F}_k$. If a three-entry sequence is given, then the first, second, and third entry is respectively $\gamma_{jk}$, $\Fop_j$, and $\Fop_k$. A \pyinline{LindbladMasterEquation} object is an instance of \pyinline{sympy.Equality}, which can be directly handled by \pyinline{CGTransform} to give the evolution of the \pyinline{StateFunction}. 

\section{Examples of Use}\label{section4}

The phase space formulation of quantum mechanics is typically done for a unipartite system; in this case, the phase space is two-dimensional, making it easy to visualize, compared to the $2N$-dimensional phase space for $N$-partite systems. In \pyinline{symqups.simple}, all phase space variables and the corresponding operators are instantiated using an empty-string subscript, i.e., the SymPy symbol \pyinline{sympy.Symbol("")}. Furthermore, the user need not instantiate the phase space variables and operators; they are directly accessible as global variables. We recommend the import statement\footnote{In Python, it is generally a bad practice to do \pyinline{from package import *} as it can pollute the namespace and risk getting overwritten by other Python objects with the same name. However, we believe that our choices of variable names are safe enough for a typical use case to justify this import paradigm.}
\begin{CodeBlock}
    from symqups.simple import *
    
    import sympy
\end{CodeBlock}
To showcase and validate the package's functionality, we consider the evolution of phase space representations of several quantum systems reported in the literature.

\subsection{The simple harmonic oscillator}

This example combines the information found in Refs.~\cite{ gardiner_2004_QuantumNoiseHandbook, mattos_2020_TimeEvolutionQuantized}. Let us consider a simple harmonic oscillator that is coherently driven and coupled to a thermal bath. We have
\begin{equation}
\begin{alignedat}{2}
    \preprintmod{
        \hat{H}=\hbar\omega\PT{ \adop\aop+\frac{1}{2}} + \PT{\lambda\adop+\lambda^*\aop}, \quad  \gamma_{11}= 2\Gamma\PT{1+n_\mathrm{avg}}, \quad  \hat{F}_1 = \aop, \quad \gamma_{22}=2\Gamma n_\mathrm{avg}, \quad \hat{F}_2=\adop,
    }{
        \hat{H}&=\hbar\omega\PT{ \adop\aop+\frac{1}{2}} + \PT{\lambda\adop+\lambda^*\aop}, 
        \\ 
        \gamma_{11}&= 2\Gamma\PT{1+n_\mathrm{avg}}, \quad  && \hat{F}_1 = \aop, 
        \\
        \gamma_{22}&=2\Gamma n_\mathrm{avg}, \quad &&\hat{F}_2=\adop,
    }
\end{alignedat}
\end{equation}
where $\omega$ is the angular frequency of the oscillator, $\lambda$ is the driving parameter, $\Gamma$ is characteristic decay constant for the system--thermal-bath coupling, and $n_\mathrm{avg}$ is the mean excitation number of the thermal bath. Setting $\Gamma=0$, the evolution of the system's $P$ representation is given by (see Section 4.5 of Ref.~\cite{gardiner_2004_QuantumNoiseHandbook})
\begin{equation}\label{eq:example_1_P_evo}
    \pdv{P}{t} = i \PT{\omega \pdv{}{\alpha}\alpha - \omega \pdv{}{\alpha^*}\alpha^* + \frac{\lambda}{\hbar} \pdv{}{\alpha} - \frac{\lambda^*}{\hbar} \pdv{}{\alpha^*}} P.
\end{equation}
If we set $\omega=\lambda=0$ instead (setting $\omega=0$ is interpreted as moving into an interaction picture), the evolution of the $Q$ function is obtained to be (see Eq. (8) of Ref.~\cite{mattos_2020_TimeEvolutionQuantized})
\begin{equation}\label{eq:example_1_Q_evo}
    \pdv{Q}{t} = \SB{\Gamma\PT{\pdv{}{\alpha}\alpha+\pdv{}{\alpha^*}\alpha^*} +2\Gamma \PT{1+n_\mathrm{avg}} \pdv{}{\alpha,\alpha^*}}Q
\end{equation}
Since this model is ubiquitous, it would be a waste to reproduce only the two equations above. Instead, let us compute the evolution for an arbitrary $s$, without making any assumptions about the physical parameters. We have 
\begin{CodeBlock}
    omega, lmbda, Gamma, n\_mean = sympy.symbols(r"omega, lambda, Gamma, n\_{\textbackslash}mathrm\{avg\}")

    H = hbar.val*omega*adOp*aOp + lmbda*adOp + dagger(lmbda*adOp)

    dissip\_1 = [2*Gamma*(1+n\_mean), aOp]

    dissip\_2 = [2*Gamma*n\_mean, adOp]

    lme = LindbladMasterEquation(H, dissip\_1, dissip\_2)

    result = collect\_by\_derivative( CGTransform(lme) )  \# see Appendix D

    result
\end{CodeBlock}
\RenderBlock{
    \preprintmod{
        \frac{\partial}{\partial t_{}} W_{- s} = 2 \Gamma W_{- s} + \left(2 \Gamma n_\mathrm{avg} + \Gamma s + \Gamma\right) \frac{\partial^{2}}{\partial \overline{\alpha}_{}\partial \alpha_{}} W_{- s} + \left(\Gamma \alpha_{} + i \omega \alpha_{} + \frac{i \lambda}{\hbar}\right) \frac{\partial}{\partial \alpha_{}} W_{- s} + \left(\Gamma \overline{\alpha}_{} - i \omega \overline{\alpha}_{} - \frac{i \overline{\lambda}}{\hbar}\right) \frac{\partial}{\partial \overline{\alpha}_{}} W_{- s}
    }{  \begin{split}
        \frac{\partial}{\partial t_{}} W_{- s} &= 2 \Gamma W_{- s} + \left(2 \Gamma n_\mathrm{avg} + \Gamma s + \Gamma\right) \frac{\partial^{2}}{\partial \overline{\alpha}_{}\partial \alpha_{}} W_{- s} 
        \\ &\quad + \left(\Gamma \alpha_{} + i \omega \alpha_{} + \frac{i \lambda}{\hbar}\right) \frac{\partial}{\partial \alpha_{}} W_{- s} 
        \\&\quad + \left(\Gamma \overline{\alpha}_{} - i \omega \overline{\alpha}_{} - \frac{i \overline{\lambda}}{\hbar}\right) \frac{\partial}{\partial \overline{\alpha}_{}} W_{- s}
        \end{split}
    }
}
For $\Gamma=0$, our result is independent of $s$. This property corresponds to the fact that ordering ambiguity is physically irrelevant for a quadratic Hamiltonian\footnote{We can straightforwardly verify that the CG bracket given in Eq.~\eqref{eq:CG_bracket} reduces to the Poisson bracket, meaning that the state function evolves identically to a classical phase space distribution under the Liouville equation. Beyond a quadratic Hamiltonian, this property does not generally hold, hence no global dynamical equivalence between the quantum and classical Hamiltonian. In canonical quantization, this obstruction is known as the Groenewold--Van Hove theorem~\cite{Groenewold1946, VanHove}.}. If we further set $\lambda=0$, then we have the phase space evolution of a quantum simple harmonic oscillator:
\begin{CodeBlock}
    result.subs({Gamma : 0, lmbda : 0})
\end{CodeBlock}
\RenderBlock{
    \frac{\partial}{\partial t_{}} \left(\pi^{} P\right) = i \omega \alpha_{} \frac{\partial}{\partial \alpha_{}} \left(\pi^{} P\right) - i \omega \overline{\alpha}_{} \frac{\partial}{\partial \overline{\alpha}_{}} \left(\pi^{} P\right)
}
Next, we set only $\Gamma=0$ and substitute $s=-1$ to obtain
\begin{CodeBlock}
    s.val = -1
    
    collect\_by\_derivative( sympy.expand(CGTransform(lme)) ).subs(Gamma, 0)
\end{CodeBlock}
\RenderBlock{
    \preprintmod{
        \frac{\partial}{\partial t_{}} \left(\pi^{} P\right) = \left(i \omega \alpha_{} + \frac{i \lambda}{\hbar}\right) \frac{\partial}{\partial \alpha_{}} \left(\pi^{} P\right) + \left(- i \omega \overline{\alpha}_{} - \frac{i \overline{\lambda}}{\hbar}\right) \frac{\partial}{\partial \overline{\alpha}_{}} \left(\pi^{} P\right)
    }{
            \begin{split}
    \frac{\partial}{\partial t_{}} \left(\pi^{} P\right) &= \left(i \omega \alpha_{} + \frac{i \lambda}{\hbar}\right) \frac{\partial}{\partial \alpha_{}} \left(\pi^{} P\right) \\&\quad + \left(- i \omega \overline{\alpha}_{} - \frac{i \overline{\lambda}}{\hbar}\right) \frac{\partial}{\partial \overline{\alpha}_{}} \left(\pi^{} P\right)
\end{split}
    }
}
which agrees with Eq.~\eqref{eq:example_1_P_evo}. Lastly, we keep $\Gamma$ as is, then set $\omega=\lambda=0$ and $s=1$ to obtain
\begin{CodeBlock}
    s.val = 1
    
    collect\_by\_derivative( sympy.expand(CGTransform(lme)) ).subs({omega:0, lmbda:0})
\end{CodeBlock}
\RenderBlock{
    \preprintmod{
        \frac{\partial}{\partial t_{}} Q = \Gamma \alpha_{} \frac{\partial}{\partial \alpha_{}} Q + \Gamma \overline{\alpha}_{} \frac{\partial}{\partial \overline{\alpha}_{}} Q + 2 \Gamma Q + \left(2 \Gamma n_\mathrm{avg} + 2 \Gamma\right) \frac{\partial^{2}}{\partial \overline{\alpha}_{}\partial \alpha_{}} Q
    }{
        \begin{split}
            \frac{\partial}{\partial t_{}} Q &= \Gamma \alpha_{} \frac{\partial}{\partial \alpha_{}} Q + \Gamma \overline{\alpha}_{} \frac{\partial}{\partial \overline{\alpha}_{}} Q + 2 \Gamma Q \\&\quad + \left(2 \Gamma n_\mathrm{avg} + 2 \Gamma\right) \frac{\partial^{2}}{\partial \overline{\alpha}_{}\partial \alpha_{}} Q
        \end{split}
    }
}
which agrees with Eq.~\eqref{eq:example_1_Q_evo}.

\subsection{The Kerr nonlinearity}

One prime example of a nonlinearity we can introduce to a simple harmonic oscillator is the Kerr nonlinearity. Despite its simplicity, it is an ubiquituous model in quantum optics to model some four wave mixing processes. In the frame rotating at the simple harmonic frequency, a system with Kerr nonlinearity is described by the Hamiltonian (with $\hbar=1$)
\begin{equation}
    \hat{H} = \frac{\kappa}{2}\PT{\adop}^2\aop^2.
\end{equation} Ref.~\cite{Propp2023} shows that the corresponding evolution of the Wigner function is given by [see Eq. (5)]
\begin{equation}
\begin{split}
    \preprintmod{
        \partial_t W = -i\kappa \PT{\alpha^*\alpha-1}\PT{\alpha^* \pdv{W}{\alpha^*} - \alpha\pdv{W}{\alpha} }
    - i\frac{\kappa}{4} \PT{\alpha \pdv[2,1]{W}{\alpha,\alpha^*} - \alpha^* \pdv[1,2]{W}{\alpha,\alpha^*}}.
    }{
        \partial_t W &= -i\kappa \PT{\alpha^*\alpha-1}\PT{\alpha^* \pdv{W}{\alpha^*} - \alpha\pdv{W}{\alpha} }
    \\&\quad - i\frac{\kappa}{4} \PT{\alpha \pdv[2,1]{W}{\alpha,\alpha^*} - \alpha^* \pdv[1,2]{W}{\alpha,\alpha^*}}.
    }
\end{split}
\end{equation}
Here we reproduce this equation using \symqups:
\begin{CodeBlock}
    kappa = sp.symbols("kappa", real=True)

    hbar.val = 1
    
    s.val = 0

    H = kappa/2 * adOp**2 * aOp**2

    lme = LindbladMasterEquation(H)

    collect\_by\_derivative( sp.expand(CGTransform(lme)) )
\end{CodeBlock}
\RenderBlock{
        \preprintmod{
        \frac{\partial}{\partial t_{}} W = - \frac{i \kappa \alpha_{} \frac{\partial^{3}}{\partial \overline{\alpha}_{}\partial \alpha_{}^{2}} W}{4} + \frac{i \kappa \overline{\alpha}_{} \frac{\partial^{3}}{\partial \overline{\alpha}_{}^{2}\partial \alpha_{}} W}{4} + \left(- i \kappa \alpha_{} \overline{\alpha}_{}^{2} + i \kappa \overline{\alpha}_{}\right) \frac{\partial}{\partial \overline{\alpha}_{}} W + \left(i \kappa \alpha_{}^{2} \overline{\alpha}_{} - i \kappa \alpha_{}\right) \frac{\partial}{\partial \alpha_{}} W
        }{
                \begin{split}
            \frac{\partial}{\partial t_{}} W &= - \frac{i \kappa \alpha_{} \frac{\partial^{3}}{\partial \overline{\alpha}_{}\partial \alpha_{}^{2}} W}{4} + \frac{i \kappa \overline{\alpha}_{} \frac{\partial^{3}}{\partial \overline{\alpha}_{}^{2}\partial \alpha_{}} W}{4} \\&\quad + \left(- i \kappa \alpha_{} \overline{\alpha}_{}^{2} + i \kappa \overline{\alpha}_{}\right) \frac{\partial}{\partial \overline{\alpha}_{}} W \\&\quad + \left(i \kappa \alpha_{}^{2} \overline{\alpha}_{} - i \kappa \alpha_{}\right) \frac{\partial}{\partial \alpha_{}} W
        \end{split}}
}

\subsection{The Stuart-Landau oscillator}

A self-sustained oscillator is an oscillator that can maintain its own rhythm thanks to having balanced sources of gain and loss, with at least one source behaving nonlinearly. At steady state, in the phase space, the phase point of a self-sustained oscillator tracks a fixed orbit known as its ``limit cycle''~\cite{Pikovsky_Rosenblum_Kurths_2001, strogatz2018nonlinear}. Recently, the quantum versions of self-sustained oscillators have gained some interest within the community. 

Of particular relevance to this work is Ref.~\cite{lim_transient_2025}, which discusses the Wigner transform of the quantum Stuart-Landau oscillator, a paradigmatic model whose classical limit cycle is circular. Setting $\hbar=\omega_0=1$, the Hamiltonian is given by $\hat{H}=\adop\aop+1/2$. Meanwhile, using the notation $\PT{\gamma_{jj}, \hat{F}_j=\hat{F}_k}$, the dissipators are given by: (1) $\PT{\kappa_1, \adop}$ which describes one-quantum gain; (2) $\PT{\gamma_1,\aop}$ describing one-quantum loss; and (3) $\PT{\gamma_2,\aop^2}$ describing two-quantum loss, which is the nonlinear process needed for the self-sustained oscillation. The corresponding equation for the Wigner function is given by $\pdv{W}/{t}=\Theta W$, where \{see Eq. (15) of Ref.~\cite{lim_transient_2025}\}
\begin{equation}
    \preprintmod{
        \begin{split}
            \Theta &= -\partial_q p + \partial_p q
            \\
            &\quad -\frac{1}{2} \PT{\kappa_1-\gamma_1} \PT{\partial_q q+\partial_p p} + \frac{1}{4}\PT{\kappa_1+\gamma_1}\PT{\partial_q^2+\partial_p^2}
            \\
            &\quad +\frac{\gamma_2}{2} \CB{
                \partial_q \SB{\PT{q^2+p^2-2}q}
                +
                \partial_p \SB{\PT{q^2+p^2-2}p}
                +
                \PT{\partial_q^2+\partial_p^2}\PT{q^2+p^2-1}
                +
                \frac{1}{4} \PT{\partial_q^3q + \partial_q^2\partial_p p + \partial_q\partial_p^2 q + \partial_p^3 p}
            }
        \end{split}
    }{ 
        \begin{split}
            \Theta &= -\partial_q p + \partial_p q
            \\
            &\quad -\frac{1}{2} \PT{\kappa_1-\gamma_1} \PT{\partial_q q+\partial_p p} + \frac{1}{4}\PT{\kappa_1+\gamma_1}\PT{\partial_q^2+\partial_p^2}
            \\
            &\quad +\frac{\gamma_2}{2} \Bigg\{
                \partial_q \SB{\PT{q^2+p^2-2}q}
                +
                \partial_p \SB{\PT{q^2+p^2-2}p}
                \\&\qquad\qquad +
                \PT{\partial_q^2+\partial_p^2}\PT{q^2+p^2-1}
                \\&\qquad\qquad +
                \frac{1}{4} \PT{\partial_q^3q + \partial_q^2\partial_p p + \partial_q\partial_p^2 q + \partial_p^3 p}
            \Bigg\} 
        \end{split}
    }
\end{equation}
in the $(q,p)$ coordinate system. It is straightforward to transform $\Theta$ into the $\alphacoord$ system using the chain rule. Following Ref.~\cite{lim_transient_2025}, we use $\alpha=\PT{q+ip}/\sqrt{2}$ to obtain 
\begin{equation}
\preprintmod{
    \begin{split}
        \Theta &= i\alpha \partial_\alpha - i\alpha^*\partial_{\alpha^*}
        \\
        &\quad 
        - \frac{1}{2} \PT{\kappa_1-\gamma_1} \PT{\partial_\alpha \alpha + \partial_{\alpha^*}\alpha^*} + \frac{1}{2} \PT{\kappa_1+\gamma_1} \partial_\alpha \partial_{\alpha^*}
        \\
        &\quad 
        +\frac{\gamma_2}{4} \CB{
            \partial_\alpha\SB{4\alpha \PT{\alpha\alpha^* - 1}}
            +
             \partial_{\alpha^*} \SB{4\alpha^*\PT{\alpha\alpha^* - 1}}
             +
            \partial_\alpha \partial_{\alpha^*} \PT{8\alpha \alpha^*-4}
            +
             \partial_\alpha^2 \partial_{\alpha^*} \alpha
            +
             \partial_\alpha \partial_{\alpha^*}^2 \alpha^*
        }
    \end{split}
}{
    \begin{split}
        \Theta &= i\alpha \partial_\alpha - i\alpha^*\partial_{\alpha^*}
        \\
        &\quad 
        - \frac{1}{2} \PT{\kappa_1-\gamma_1} \PT{\partial_\alpha \alpha + \partial_{\alpha^*}\alpha^*} + \frac{1}{2} \PT{\kappa_1+\gamma_1} \partial_\alpha \partial_{\alpha^*}
        \\
        &\quad 
        +\frac{\gamma_2}{4} \Bigg\{
            \partial_\alpha\SB{4\alpha \PT{\alpha\alpha^* - 1}}
            +
             \partial_{\alpha^*} \SB{4\alpha^*\PT{\alpha\alpha^* - 1}}
             \\&\qquad\qquad +
            \partial_\alpha \partial_{\alpha^*} \PT{8\alpha \alpha^*-4}
            \\&\qquad\qquad +
             \partial_\alpha^2 \partial_{\alpha^*} \alpha
            +
             \partial_\alpha \partial_{\alpha^*}^2 \alpha^*
        \Bigg\}
    \end{split}
}
\end{equation}
An equivalent result can be obtained using \symqups:
\begin{CodeBlock}
    kappa\_1, gamma\_1, gamma\_2 = sp.symbols("kappa\_1, gamma\_1, gamma\_2", real=True)

    hbar.val = 1
    
    s.val = 0

    H = adOp*aOp + sp.Rational(1,2)

    dissip\_1 = [kappa\_1, adOp]

    dissip\_2 = [gamma\_1, aOp]

    dissip\_3 = [gamma\_2, aOp**2]

    lme = LindbladMasterEquation(H, dissip\_1, dissip\_2, dissip\_3)

    collect\_by\_derivative( sp.expand(CGTransform(lme)) )
\end{CodeBlock}
\RenderBlock{
    \preprintmod{
        \frac{\partial}{\partial t_{}} W &= \gamma_{1} W + 4 \gamma_{2} \alpha_{} \overline{\alpha}_{} W + \frac{\gamma_{2} \alpha_{} \frac{\partial^{3}}{\partial \overline{\alpha}_{}\partial \alpha_{}^{2}} W}{4} + \frac{\gamma_{2} \overline{\alpha}_{} \frac{\partial^{3}}{\partial \overline{\alpha}_{}^{2}\partial \alpha_{}} W}{4} - \kappa_{1} W 
        + \left(\frac{\gamma_{1}}{2} + 2 \gamma_{2} \alpha_{} \overline{\alpha}_{} + \frac{\kappa_{1}}{2}\right) \frac{\partial^{2}}{\partial \overline{\alpha}_{}\partial \alpha_{}} W 
        \\
        &\quad + \left(\frac{\gamma_{1} \alpha_{}}{2} + \gamma_{2} \alpha_{}^{2} \overline{\alpha}_{} + \gamma_{2} \alpha_{} - \frac{\kappa_{1} \alpha_{}}{2} + i \alpha_{}\right) \frac{\partial}{\partial \alpha_{}} W + \left(\frac{\gamma_{1} \overline{\alpha}_{}}{2} + \gamma_{2} \alpha_{} \overline{\alpha}_{}^{2} + \gamma_{2} \overline{\alpha}_{} - \frac{\kappa_{1} \overline{\alpha}_{}}{2} - i \overline{\alpha}_{}\right) \frac{\partial}{\partial \overline{\alpha}_{}} W
    }{
        \begin{split}
            \frac{\partial}{\partial t_{}} W &= \gamma_{1} W + 4 \gamma_{2} \alpha_{} \overline{\alpha}_{} W + \frac{\gamma_{2} \alpha_{} \frac{\partial^{3}}{\partial \overline{\alpha}_{}\partial \alpha_{}^{2}} W}{4} \\&\quad + \frac{\gamma_{2} \overline{\alpha}_{} \frac{\partial^{3}}{\partial \overline{\alpha}_{}^{2}\partial \alpha_{}} W}{4} - \kappa_{1} W 
            \\
            &\quad + \left(\frac{\gamma_{1}}{2} + 2 \gamma_{2} \alpha_{} \overline{\alpha}_{} + \frac{\kappa_{1}}{2}\right) \frac{\partial^{2}}{\partial \overline{\alpha}_{}\partial \alpha_{}} W 
             \\
            &\quad + \left(\frac{\gamma_{1} \alpha_{}}{2} + \gamma_{2} \alpha_{}^{2} \overline{\alpha}_{} + \gamma_{2} \alpha_{} - \frac{\kappa_{1} \alpha_{}}{2} + i \alpha_{}\right) \frac{\partial}{\partial \alpha_{}} W \\
            &\quad + \left(\frac{\gamma_{1} \overline{\alpha}_{}}{2} + \gamma_{2} \alpha_{} \overline{\alpha}_{}^{2} + \gamma_{2} \overline{\alpha}_{} - \frac{\kappa_{1} \overline{\alpha}_{}}{2} - i \overline{\alpha}_{}\right) \frac{\partial}{\partial \overline{\alpha}_{}} W
        \end{split}
    }
}


\section{Performance}\label{section5}


\begin{figure}[!t]
    \centering
    \includegraphics[width=\preprintmod{0.75}{1}\linewidth]{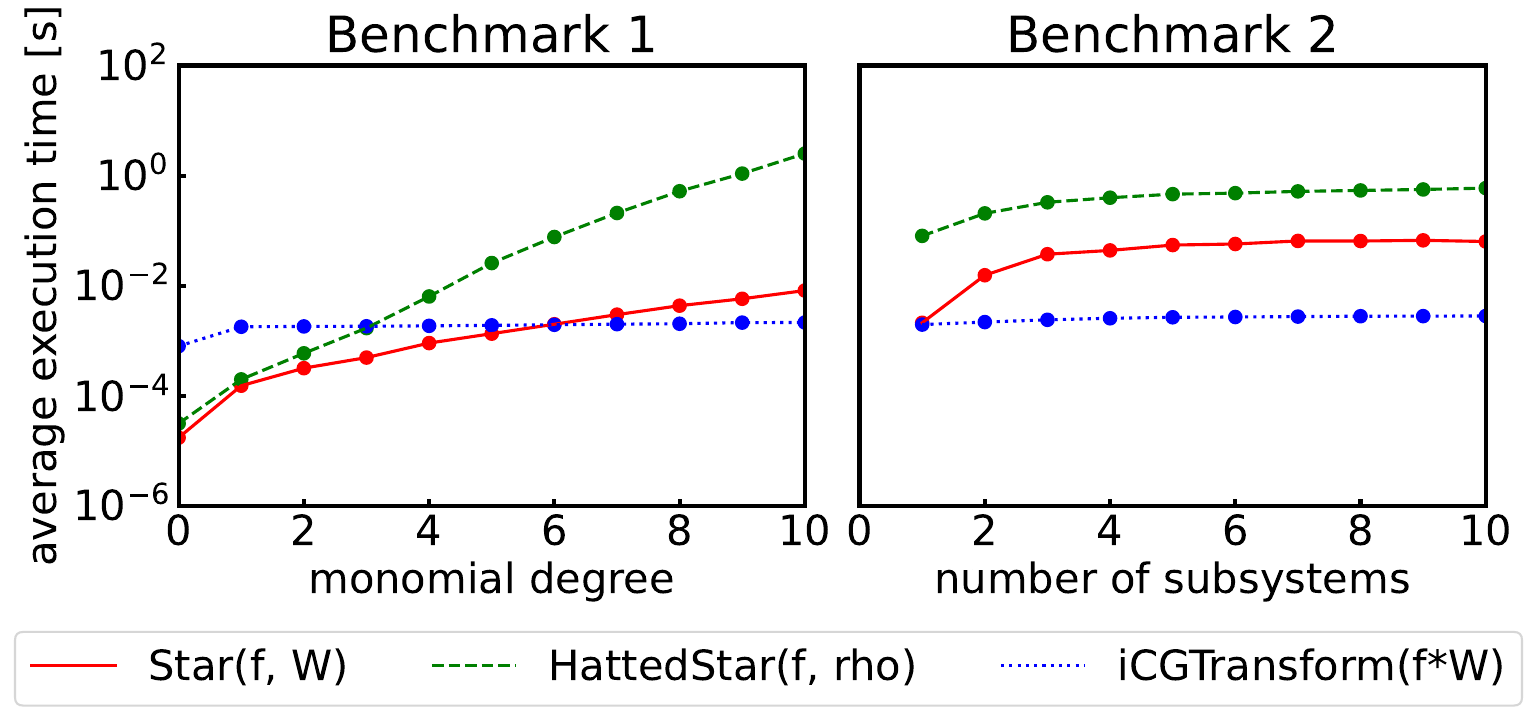}
    \caption{Performance benchmark for \symqups. We record the average time it takes to run the functions indicated in the figure legend $100$ times. Here, \pyinline{f} is a random monomial in $\aopadop$ for \pyinline{HattedStar}, and $\alphacoord$ for \pyinline{Star} and \pyinline{iCGTransform}, with unit coefficient. In ``Benchmark 1'', the degree of the monomial is varied and only one subsystem is considered. In ``Benchmark 2'', the degree of monomial is fixed at $6$, while the number of subsystems is varied. The random monomials are generated using \pyinline{get\_random\_poly}, which does not support randomizer seeds, so repeated benchmarks may slightly differ from each other. 
    }
    \label{fig:benchmark}
\end{figure}

Like SymPy on which it is based, \symqups~relies heavily on recursion. Since the mappings implemented by the package are linear, \symqups~assumes that an input takes the general form of a sum of terms, i.e., a \pyinline{sympy.Add} object. To ensure that an input conforms to this general form, \symqups~applies preprocessing to function arguments, which calls SymPy's \pyinline{.doit}, followed by \pyinline{.expand} method. For such an input, the mappings are sequentially applied to each summand, thereby forming the top-level for-loop with linear time complexity. Additionally, the extra overheads from the preprocessing may become noticable for more complex inputs.

Each summand is generally treated as a \pyinline{sympy.Mul} object, which is where the nontriviality of the implementations lie. For example, the \pyinline{CGTransform} of such an object is given by the \pyinline{Star} product of the \pyinline{CGTransform} of its factors, i.e., objects contained in its \pyinline{.args} attribute. However, the straightforward algorithm based on this defining property might not be the most efficient. For instance, if the input to \pyinline{CGTransform} is simply an \pyinline{sOrdering} bracket, then the package can instead return the operator-to-scalar substitution of the bracket's \pyinline{.content}, using the function \pyinline{op2sc}. Various \emph{evaluation shortcuts} we apply results in a somewhat complex decision tree in each class constructor, in exchange for faster evaluations for some special input forms. 

We perform benchmarks on \pyinline{Star}, \pyinline{HattedStar}, and \pyinline{iCGTransform}, but not on \pyinline{CGTransform} since it internally relies on \pyinline{Star} to evaluate products. The benchmarks are run on a 2025 MacBook Air equipped with Apple M4 chip with 16 GB of unified memory, running on the operating system macOS Tahoe 26.5. The code is run on Visual Studio Code version 1.127.0, with Python 3.14.4 running SymPy 1.14.0. The benchmark results are shown in Fig.~\ref{fig:benchmark}. 

For our first benchmark, we consider a unipartite system and take an average over $100$ \pyinline{Star} products between a random mononomial in $\alphacoord$ with $W_s$, recording the change in average execution time as we increase the monomial degree. The benchmark is similar for the \pyinline{HattedStar} products, with $\alphacoord$ replaced by $\aopadop$ and $W_s$ replaced by $\rho$. Meanwhile, a random monomial in $\alphacoord$ times $W$ is input into \pyinline{iCGTransform}. The execution of all functions takes longer to evaluate the functions as the monomial degree is increased. The average runtime for \pyinline{HattedStar} quickly outgrows the other two functions, due to the noncommutativity of operators forcing us to loop through each operator. As a comparison, for \pyinline{Star}, we can treat powers of the scalar variables in a single loop iteration. Meanwhile, the average runtime for \pyinline{iCGTransform}  increases slowly, a behavior we expect from implementing Eq.~\eqref{eq:hsbs_applied_to_iCG} using a two-level nested loop over the powers of $\alpha$ and $\alpha^*$.

In the second benchmark, we fix the monomial degree at $6$ and record the average execution time as we increase the number of subsystems. As the number of subsystems is increased from $1$, we see a growth in average execution time which quickly slows down. This growth is more pronounced for \pyinline{Star}, since for multipartite systems \pyinline{f} contains more arguments through which \pyinline{Star} iterates. Curiously, the number of objects for \pyinline{HattedStar} to iterate through only depends on the monomial order, so we can expect the average runtime to stay constant, which is not the case. We may associate this behavior with \pyinline{Derivative} having more variables to evaluate. Meanwhile, the average runtime for \pyinline{iCGTransform} stays pretty much constant.

Setting time complexity aside, it is unlikely that a typical usage of the package in quantum mechanics involves monomial degree and number of subsystems larger than our benchmark domain presented here. In the worst case, it may only take several minutes of computation time using a modern hardware, which is a significant speedup compared to conventional evaluations. Adding to the fact that SymPy is single-threaded, we conclude that \symqups~is \emph{computationally practical}.


\section{Summary and Outlook}\label{section6}

We have developed a Python symbolic package \symqups~for efficient quantum phase space algebra (and, by extension, quantization). Out of the various functionalities it offers (see Appendix~\ref{appsec:docs}), perhaps the most important one is the evaluation of the Cahill-Glauber (which includes Glauber-Sudarshan $P$, Wigner, and Husimi $Q$) transform of the widely-used Lindblad master equation. We have described the core functionalities of the package and demonstrated its capability in reproducing results within the literature, the latter of which also serves to validate its workings. As shown by our performance testing, it barely requires any time to perform theoretically relevant computations with a typical modern hardware, allowing it to offer significant speedup over traditional computation methods.

Even in this era of artifical intelligence (AI) models that exhibit remarkable capabilities in solving complicated problems, they remain statistical models rather than symbolic computation engines. Their outputs become increasingly unreliable for lengthy symbolic manipulations, a limitation that is especially prominent in the phase space formulation of quantum mechanics, whose niche nature provides a relatively small pool of training data. A ``calculator'' like \symqups~thus remains desirable, as it completely eliminates the uncertainties inherent in AI-based computations.  We welcome suggestions and constructive criticism to improve this package at \url{https://github.com/hendry24/SymQuPS/issues}.

\medskip
\noindent\textbf{Declaration of competing interest}
\medskip

The authors declare that they have no known competing financial interests or personal relationships that could have appeared to influence the work reported in this paper.

\medskip
\noindent\textbf{Data availability}
\medskip

The package's source code and instructions for installation are publicly available at \url{https://github.com/hendry24/SymQuPS}. The code used for Section~\ref{section4} is compiled into a {Jupyter Notebook} available at \url{https://github.com/hendry24/SymQuPS/blob/main/paper/examples-of-use.ipynb}. The code used for Section~\ref{section5} is compiled into a {Jupyter Notebook} available at \url{https://github.com/hendry24/SymQuPS/blob/main/paper/benchmark.ipynb}. Online documentation is available at \url{https://symqups.readthedocs.io/}.


\appendix

\section{Multipartite equivalent of relevant expressions}\label{appsec:multipartite}

For an $N$-partite system, let $\bm{\alpha}=\PT{\alpha_0,\dots,\alpha_{N-1}}^\intercal$. The multipartite Cahill-Glauber (CG) transform of some Hilbert space operator $\hat{F}$ is given by
\begin{equation}
    f^{(s)}\PT{\bm{\alpha}, \bm{\alpha}^*} = \CG{\hat{F}}\PT{\bm{\alpha}, \bm{\alpha}^*} = \mathrm{Tr}\PT{\hat{F}\bigotimes_{j=0}^{N-1}{\mathcal{T}}_{-s}^{(j)}},
\end{equation}
where the $j$th subsystem's CG kernel is given by
\begin{equation}\label{eq:CG_kernel_multipartite}
\preprintmod{
    \exp\PT{\frac{s}{2}\VB{\beta}^2 + \alpha_j\beta^* - \alpha^*_j\beta} , 
    \quad
    \hat{D}_j\PT{\beta}=e^{\beta\adop_j-\beta^*\aop_j}.
}{
\begin{split}
    \mathcal{T}_{s}^{(j)}\PT{\alpha_j,\alpha_j^*} &= \int_{\mathbb{R}^2} \frac{\odif[order=2]{\beta}}{\pi} \hat{D}_j\PT{\beta} \\
    &\qquad \qquad\times \exp\PT{\frac{s}{2}\VB{\beta}^2 + \alpha_j\beta^* - \alpha^*_j\beta} , 
\end{split}
}
\end{equation}
\preprintmod{}{where $\hat{D}_j\PT{\beta}=e^{\beta\adop_j-\beta^*\aop_j}$}. The inverse Cahill-Glauber (iCG) transform of some phase space function $f=f\PT{\bm{\alpha}, \bm{\alpha}^*}$ is given by
\begin{equation}
    \hat{F}^{(s)} = \iCG{f} = \int_{\mathbb{R}^{2N}} \frac{\odif[order=2]{\alpha_0} \dots \odif[order=2]{\alpha_{N-1}}}{\pi^N} f \bigotimes_{j=0}^{N-1} \mathcal{T}_s^{(j)} .
\end{equation}
Next, the $N$-partite star product is given by
\begin{equation}
    \Star = \exp\PT{\sum_{j=0}^{N-1} \SB{\frac{s+1}{2} \overset{\leftarrow}{\partial}_{\alpha_j} \overset{\rightarrow}{\partial}_{\alpha^*_j} + \frac{s-1}{2} \overset{\leftarrow}{\partial}_{\alpha^*_j} \overset{\rightarrow}{\partial}_{\alpha_j}}}
\end{equation}
Lastly, the $N$-partite hatted star product is given by
\begin{equation}
    \dStar = \exp\PT{\sum_{j=0}^{N-1} \SB{-\frac{s+1}{2}\overset{\leftarrow}{\partial}_{\aop_j} \overset{\rightarrow}{\partial}_{\adop_j}-\frac{s-1}{2}\overset{\leftarrow}{\partial}_{\adop_j} \overset{\rightarrow}{\partial}_{\aop_j}}}.
\end{equation}

\section{Explicit formulae for core functionalities}\label{appsec:explicit_formulae}

\noindent \textbf{\pyinline{Star} product between a monomial and some test function, using the Bopp shift.} This case is straightforward. We have
\begin{equation}
\preprintmod{
    \begin{split}
    \left(\alpha^*\right)^m\alpha^n \Star g\alphacoord &= \left(\alpha^* + \frac{s-1}{2}\pdv{}{\beta}\right)^m \left(\alpha + \frac{s+1}{2}\pdv{}{\beta^*}\right)^n g\left(\beta,\beta^*\right) \Bigg|_{\beta\to\alpha}
    \\
    &= \sum_{j=0}^{m}\sum_{k=0}^{n} \binom{m}{j} \binom{n}{k} \left(\alpha^*\right)^{m-j}\alpha^{n-k}\left(\frac{s\textcolor{red}{-}1}{2}\right)^{j} \left(\frac{s\textcolor{red}{+}1}{2}\right)^{k} \pdv[j,k]{g}{\alpha,\left(\alpha^*\right)},
\end{split}
}{
    \begin{split}
    &\left(\alpha^*\right)^m\alpha^n \Star g\alphacoord 
    \\
    &\quad =\left(\alpha^* + \frac{s-1}{2}\pdv{}{\beta}\right)^m \left(\alpha + \frac{s+1}{2}\pdv{}{\beta^*}\right)^n g\left(\beta,\beta^*\right) \Bigg|_{\beta\to\alpha} 
    \\
    &\quad =\sum_{j=0}^{m}\sum_{k=0}^{n} \binom{m}{j} \binom{n}{k} \left(\alpha^*\right)^{m-j}\alpha^{n-k}
    \\
    &\qquad\qquad\quad  \times \left(\frac{s\textcolor{red}{-}1}{2}\right)^{j} \left(\frac{s\textcolor{red}{+}1}{2}\right)^{k} \pdv[j,k]{g}{\alpha,\left(\alpha^*\right)},
\end{split}
}
\end{equation}
and
\begin{equation}
\preprintmod{
    \begin{split}
    f\alphacoord \Star \left(\alpha^*\right)^m\alpha^n &= \left(\alpha^* + \frac{s+1}{2}\pdv{}{\beta}\right)^m \left(\alpha + \frac{s-1}{2}\pdv{}{\beta^*}\right)^n f\left(\beta,\beta^*\right) \Bigg|_{\beta\to\alpha}
    \\
    &= \sum_{j=0}^{m}\sum_{k=0}^{n} \binom{m}{j} \binom{n}{k} \left(\alpha^*\right)^{m-j}\alpha^{n-k}\left(\frac{s \textcolor{red}{+}1}{2}\right)^{j} \left(\frac{s\textcolor{red}{-}1}{2}\right)^{k} \pdv[j,k]{f}{\alpha,\left(\alpha^*\right)}.
\end{split}
}{
\begin{split}
    &f\alphacoord \Star \left(\alpha^*\right)^m\alpha^n 
    \\&\quad = \left(\alpha^* + \frac{s+1}{2}\pdv{}{\beta}\right)^m \left(\alpha + \frac{s-1}{2}\pdv{}{\beta^*}\right)^n f\left(\beta,\beta^*\right) \Bigg|_{\beta\to\alpha}
    \\
    &\quad = \sum_{j=0}^{m}\sum_{k=0}^{n} \binom{m}{j} \binom{n}{k} \left(\alpha^*\right)^{m-j}\alpha^{n-k}
    \\
    &\qquad\qquad\quad\times \left(\frac{s \textcolor{red}{+}1}{2}\right)^{j} \left(\frac{s\textcolor{red}{-}1}{2}\right)^{k} \pdv[j,k]{f}{\alpha,\left(\alpha^*\right)}.
\end{split}
}
\end{equation}
The generalization to the multipartite case is straightforward. 

\noindent \textbf{\pyinline{HattedStar} product between a monomial and some test function, using the Bopp shift.} We can write a polynomial in $\aop$ and $\adop$ in any ordering of our choice, so the term ``monomial'' here best refers to a ``boson string'' (a term used prolifically by Blasiak, e.g., in Ref.~\cite{blasiak_combinatorics_2005}). Let
$\prod_{j=N-1}^0 \hat{x}_j = \hat{x}_{N-1}\dotsm\hat{x}_{1}\hat{x}_0;\, \hat{x}_j\in \left\{\aop,\adop\right\}$ be a boson string of order $(m,n)$ where $m$ is the number of $\adop$ and $n$ is the number of $\aop$. We have the Bopp shifts
\begin{subequations}
\begin{align}
    \hat{x}_j \dStar \Gop\aopadop &= \hat{x}_j\Gop\aopadop + \xi_j^R \pdv{\Gop\left(\hat{b},\hat{b}^\dagger\right)}{\hat{y}_j^\dagger}\Bigg|_{\hat{b}\to\aop},
    \\
    \Gop\aopadop \dStar \hat{x}_j &= \Gop\aopadop \hat{x}_j + \xi_j^L \pdv{\Gop\left(\hat{b},\hat{b}^\dagger\right)}{\hat{y}_j^\dagger}\Bigg|_{\hat{b}\to\aop},
\end{align}
\end{subequations}
where
\begin{subequations}
\begin{align}
    \hat{y}_j &=  \hat{x}_j\Big|_{\aop\to\hat{b}},
    \\
    \xi_j^d &= \begin{cases}\displaystyle
        -\frac{s+1}{2}& \text{if $\left(\hat{x}_j,d\right)=\left(\hat{a},R\right), \left(\hat{a}^\dagger, L\right)$},
        \\ 
        \displaystyle
        -\frac{s-1}{2} & \text{if $\left(\hat{x}_j,d\right)=\left(\hat{a}^\dagger,R\right), \left(\hat{a}, L\right)$},
    \end{cases}
\end{align}
\end{subequations}
from which we can work out that
\begin{equation}
\preprintmod{
    \begin{split}
    \left(\prod_{j=N-1}^0 \hat{x}_j\right) \dStar \Gop\aopadop = 
    \sum_{w\in \CB{0,1}^N} \PT{\prod_{\substack{j=N-1,\\w_j=0}}^0 \hat{x}_j} \PT{\prod_{\substack{j=N-1,\\w_j=1}}^0 \xi_j^R \pdv*{}{\hat{x}_j^\dagger}}\hat{G}\aopadop,
\end{split}
}{
    \begin{split}
    &\left(\prod_{j=N-1}^0 \hat{x}_j\right) \dStar \Gop\aopadop \\&\quad = 
    \sum_{w\in \CB{0,1}^N} \PT{\prod_{\substack{j=N-1,\\w_j=0}}^0 \hat{x}_j} \PT{\prod_{\substack{j=N-1,\\w_j=1}}^0 \xi_j^R \pdv*{}{\hat{x}_j^\dagger}}\hat{G}\aopadop,
\end{split}
}
\end{equation}
and
\begin{equation}
\preprintmod{
\begin{split}
    \Gop\aopadop \dStar \PT{\prod_{j=N-1}^0\hat{x}_j} &=
     \sum_{w\in\CB{0,1}^N} \PT{\prod_{\substack{j=N-1,\\w_j=1}}^0 \xi_j^L \pdv*{}{\hat{x}_j^\dagger}} \Gop\aopadop \PT{\prod_{\substack{j=N-1,\\w_j=0}}^0 \hat{x}_j}.
\end{split}
}{
\begin{split}
    &\Gop\aopadop \dStar \PT{\prod_{j=N-1}^0\hat{x}_j} \\&\quad =
     \sum_{w\in\CB{0,1}^N} \PT{\prod_{\substack{j=N-1,\\w_j=1}}^0 \xi_j^L \pdv*{}{\hat{x}_j^\dagger}} \Gop\aopadop \PT{\prod_{\substack{j=N-1,\\w_j=0}}^0 \hat{x}_j}.
\end{split}
}
\end{equation}
The sums run through all words $w=\PT{w_{N-1},\dots,w_0}$ made of the alphabet $\CB{0,1}$, whose letters $w_j$ indicate whether to choose the operator $\hat{x}_j$ or its Bopp shift for the given term labeled by $w$. For the $N$-partite case, we have $\hat{x}_j \in \CB{\aop_{N-1},\adop_{N-1},\dots,\aop_0,\adop_0}$. 

\noindent \textbf{\pyinline{iCGTransform} of a product.} A generic phase space expression can be written as $\PT{\alpha^*}^m \alpha^n f\alphacoord$, where $f$ is the nonpolynomial factor. Using the HSBSs given in Eq.~\eqref{eq:HSBS}, we have
\begin{equation}\label{eq:hsbs_applied_to_iCG}
\preprintmod{
\iCG{\PT{\alpha^*}^m \alpha^n f} = \sum_{j=0}^m \sum_{k=0}^n \binom{m}{j}\binom{n}{k}  \PT{\frac{1+s}{2}}^{m-j+k} \PT{\frac{1-s}{2}}^{n+j-k} \PT{\adop}^{m-j} \aop^{n-k} \hat{F}^{(s)} \aop^k \PT{\adop}^j.
}{
\begin{split}
    &\iCG{\PT{\alpha^*}^m \alpha^n f} \\&\quad = \sum_{j=0}^m \sum_{k=0}^n \binom{m}{j}\binom{n}{k}  \PT{\frac{1+s}{2}}^{m-j+k} \PT{\frac{1-s}{2}}^{n+j-k} \\&\qquad\qquad\quad\times \PT{\adop}^{m-j} \aop^{n-k} \hat{F}^{(s)} \aop^k \PT{\adop}^j.
\end{split}
}
\end{equation}
The generalization for multipartite cases is straightforward: we can evaluate $\iCG{\PT{\alpha_2^*}^{m_2} \alpha_2^{n_2} \PT{\alpha_1^*}^{m_1} \alpha_1^{n_1} f}$ by evaluating $\iCG{\PT{\alpha_2^*}^{m_2} \alpha_2^{n_2} g}$ with $g=\PT{\alpha_1^*}^{m_1} \alpha_1^{n_1} f$, i.e., composing Eq.~\eqref{eq:hsbs_applied_to_iCG}.

\section{Other Bopp-shiftable expressions}\label{appsec:other_boppable}

Polynomials are not the only expressions for which we have a nice Bopp-shifted form. For example, we may have the function
\begin{equation}
    f(\alpha,\alpha^*) = e^{\alpha},
\end{equation}
for which a Bopp-shift evaluation is possible, say,
\begin{equation}
\preprintmod{
    f\left(\alpha,\alpha^*\right) \Star g\left(\alpha,\alpha^*\right) 
    = 
    e^{\alpha + \frac{s+1}{2} \DRdiff{\alpha^*} }
    g\left(\alpha,\alpha^*\right)
    = 
    e^\alpha g\left(\alpha,\alpha^* + \frac{s+1}{2}\right).
}{\begin{split}
    f\left(\alpha,\alpha^*\right) \Star g\left(\alpha,\alpha^*\right) 
    &= 
    e^{\alpha + \frac{s+1}{2} \DRdiff{\alpha^*} }
    g\left(\alpha,\alpha^*\right)
    \\&= 
    e^\alpha g\left(\alpha,\alpha^* + \frac{s+1}{2}\right).
    \end{split}
}
\end{equation}
Unfortunately, we could not determine a general class of such non-polynomial Bopp-able expressions, nor their relevance in actual theories. To preserve performance, then, the package does not implement special algorithms for them. We are open to adding this feature in the future. 


\section{User-level functionalities of \symqups} \label{appsec:docs}

Table~\ref{tab:docs} lists and briefly describes all user-level functionalities of \symqups.

\preprintmod{}{\clearpage \onecolumn}

\begin{table}[!t]
    \centering
    \begin{tabularx}{\textwidth}{|l|X|}
    \hline

        \pyinline{PhaseSpaceBoppOperator}
        &
        Construct a phase space Bopp operator (PSBO) or directly apply to another scalar.
        \\\hline

        \pyinline{PSBO}
        &
        Alias of \pyinline{PhaseSpaceBoppOperator}.
        \\\hline

        \pyinline{HilbertSpaceBoppSuperoperators}
        &
        Construct a Hilbert space Bopp superoperator (HSBS) or directly apply to another operator.
        \\\hline

        \pyinline{HSBS}
        &
        Alias of \pyinline{HilbertSpaceBoppSuperoperator}
        \\\hline
        
      \pyinline{CGTransform}
      &
      Return the Cahill-Glauber transform of the input.
      \\\hline
      
      \pyinline{iCGTransform}
      &
      Return the inverse Cahill-Glauber transform of the input.
      \\\hline
      \pyinline{LindbladMasterEquation}
      &
      Construct an arbitrary Lindblad master equation. 
      \\\hline
      
      \pyinline{LME}
      &
      Alias of \pyinline{LindbladMasterEquation}
      \\\hline
      
      \pyinline{alpha2qp}
      &  
      Return an equivalent expression with $\alphacoord$ written in terms of $(q,p)$, similarly for the operators.
      \\\hline
      
      \pyinline{qp2alpha}   
      &  
      The reverse of \pyinline{alpha2qp}.
      \\\hline
      
      \pyinline{sc2op}
      &
      Return a non-equvalent expression with $\PT{q,p,\alpha,\alpha^*}$ replaced by their operator counterparts.
      \\\hline
      
      \pyinline{op2sc}
      &
      The reverse of \pyinline{sc2op}.
      \\\hline
      
      \pyinline{dagger}
      &
      Return the Hermitian conjugate.
      \\\hline
      \pyinline{explicit\_sOrdering}
      &
      Return the input expression with \pyinline{sOrdering} brackets written explicitly for $s=-1,0,1$.
      \\\hline

      \pyinline{express\_sOrdering}
      &
      Return an equivalent expression with \pyinline{sOrdering} brackets expanded in terms of \pyinline{sOrdering} brackets with the specified parameter.
      \\\hline
      \pyinline{normal\_ordered\_equivalent}
      &
      Return an equivalent expression where all ladder operator monomials are normal-ordered. For example, inputting $\aop\adop$ returns $\adop\aop+1$. Uses Blasiak's explicit formula~\cite{blasiak_combinatorics_2005}, which is also implemented by Ref.~\cite{lim_2025_PyBoLaNOPythonSymbolic}.
      \\\hline
      \pyinline{s\_ordered\_equivalent}
      &
      Return an equivalent expression where all ladder operator monomials are $s$-ordered.
      \\\hline

        \pyinline{Derivative}
        &
        The package's derivative object which deals with the formal derivatives with respect to \pyinline{Operator} objects. Returns the resulting expression or a \pyinline{sympy.Derivative} object thereof.
        \\\hline

        \pyinline{Commutator}
        &
        Construct a commutator object.
        \\\hline
        
      \pyinline{normal\_order}
      &
      Return a non-equivalent expression that is the normal ordering of the input.
      \\\hline
      
      \pyinline{Weyl\_order}
      &
      Return a non-equivalent expression that is the Weyl (or symmetric) ordering of the input.
      \\\hline
      
      \pyinline{antinormal\_order}
      &
      Return a non-equivalent expression that is the antinormal ordering of the input.
      \\\hline

        \pyinline{Star}
        &
        Compute the star product.
        \\\hline

        \pyinline{HattedStar}
        &
        Compute the hatted star product.
        \\\hline
      
      \pyinline{s\_quantize}
      &
      Canonical-quantize the input using $s$-ordering. An alias to \pyinline{iCGTransform} to emphasize its role as a canonical quantization.
      \\\hline
      
      \pyinline{normal\_quantize}
      &
      Canonical-quantize the input using normal ordering.
      \\\hline
      
      \pyinline{Weyl\_quantize}
      &
      Canonical-quantize the input using Weyl (or symmetric) ordering.
      \\\hline
      \pyinline{antinormal\_quantize}
      &
      Canonical-quantize the input using antinormal ordering.
      \\\hline

        \pyinline{get\_N}
        &
        Get the dimensionality of the problem. Equals to the number of \pyinline{sub}'s specified so far in the session.
        \\\hline

      \pyinline{get\_random\_poly}
      &
      Return a random polynomial in specified variables.
      \\\hline

    \pyinline{collect\_by\_derivative}
      &
      Collect terms containing the same \pyinline{sympy.Derivative} objects.

    \\\hline
    \end{tabularx}
    \caption{Other functionalities of the package alongside brief description of each. The full documentation and usage examples are available online at
\url{https://symqups.readthedocs.io/}.}
    \label{tab:docs}
\end{table}
\preprintmod{}{\twocolumn}

\bibliographystyle{elsarticle-num}
\bibliography{bibliography}

\end{document}